\documentclass[a4paper,fleqn,final]{cas-dc}

\usepackage[T1]{fontenc}
\usepackage[utf8]{inputenc}
\usepackage[main=english]{babel}
\DeclareUnicodeCharacter{03C1}{\ensuremath{\rho}}
\DeclareUnicodeCharacter{03B2}{\ensuremath{\beta}}
\DeclareUnicodeCharacter{03BC}{\ensuremath{\mu}}
\usepackage{textcomp}
\usepackage{hyperref}
\usepackage{amsmath,amssymb}
\usepackage{graphicx}
\usepackage{booktabs}
\usepackage{tabularx}
\usepackage{array}
\usepackage{calc}

\usepackage{threeparttable}
\usepackage{longtable}
\usepackage{adjustbox}
\usepackage[numbers,sort&compress]{natbib}
\usepackage{caption}
\usepackage{etoolbox}
\usepackage{placeins}
\usepackage{float}
\usepackage{xurl}
\usepackage{microtype}
\usepackage{lastpage}

\makeatletter
\g@addto@macro{\UrlBreaks}{\do\_\do\-\do\/\do\.\do\0\do\1\do\2\do\3\do\4\do\5\do\6\do\7\do\8\do\9}
\makeatother

\renewcommand{\ttfamily}{\rmfamily}
\AtBeginDocument{%
  \urlstyle{same}%
}

\usepackage{xcolor}
\definecolor{linknavy}{RGB}{0,70,127}
\hypersetup{
  colorlinks=true,
  linkcolor=black,
  citecolor=linknavy,
  urlcolor=linknavy,
  runcolor=linknavy
}

\AtBeginEnvironment{table}{\let\sffamily\rmfamily}
\AtBeginEnvironment{figure}{\let\sffamily\rmfamily}
\AtBeginEnvironment{table*}{\let\sffamily\rmfamily}
\AtBeginEnvironment{figure*}{\let\sffamily\rmfamily}

\ExplSyntaxOn
\RenewDocumentCommand \firstname {}
  { \textcolor{black}{\seq_use:Nn \l_stm_au_seq { ~ }} }

\RenewDocumentCommand \emailauthor { m m }
   {
     \int_gincr:N \g_ead_int
     \seq_gput_right:Nn \g_stm_ead_seq
       {
         { \href{mailto:#1}{\rmfamily #1} }
         \parsename { #2 }
         \space(\eadauthor)
       }
     }

\cs_set:Npn \__first_footerline:
{
  \group_begin:
  \small
  \normalfont
  \ifnum\theblind>0\relax
  \else
  \__short_authors: :~
  \fi
  \itshape Preprint~submitted~to~Frontiers
  \group_end:
}

\RenewDocumentCommand \printorcid { } { }

\cs_set:Npn \__first_head:
{
  \parbox[t]{\textwidth}
  {
    \rule{\textwidth}{0pt}
  }
}

\cs_set:Npn \__cas_head:
{
  \parbox{\textwidth}
  {
    \rule{\textwidth}{0pt}
  }
}

\cs_set:Npn \__cas_foot:
{
  \parbox[t]{\textwidth}
  {
   \rule{\textwidth}{.2pt}\\
   \small
   \normalfont
   \__first_footerline:
   \hfill Page~\thepage {}~of~ \lastpage
  }
}
\ExplSyntaxOff

\makeatletter
\ps@cas
\makeatother

\begin{document}

% IEEE BST control: show full author names (no ------ dashes)
\makeatletter
\def\bstctlcite{\@ifnextchar[{\@bstctlcite}{\@bstctlcite[@auxout]}}
\def\@bstctlcite[#1]#2{\@bsphack
  \@for\@citeb:=#2\do{%
    \edef\@citeb{\expandafter\@firstofone\@citeb}%
    \if@filesw\immediate\write\csname #1\endcsname{\string\citation{\@citeb}}\fi}%
  \@esphack}
\makeatother
\bstctlcite{IEEEbsTcontrol}

\shortauthors{Farajpoor and Narimani}
\shorttitle{Spatially validated Palisades wildfire vulnerability}

\title[mode=title]{Explainable GeoAI for urban wildfire structure loss: vegetation and built environment in the 2025 Palisades Fire}

\author[1]{Parastoo Farajpoor}
\author[1]{Mohammadreza Narimani}
\cormark[1]
\ead{mnarimani@ucdavis.edu}

\affiliation[1]{organization={Department of Biological and Agricultural Engineering, University of California, Davis},city={Davis},state={CA},postcode={95616},country={USA}}

\cortext[1]{Corresponding author. ORCID: Mohammadreza Narimani 0009-0001-7302-405X}

\begin{abstract}
Urban wildfire resilience depends on how built form, vegetation condition, and extreme fire weather interact, yet city-scale decision tools often combine spatial data without testing whether their apparent predictive skill transfers across neighborhoods. We developed a spatially validated GeoAI and urban-informatics workflow for the January 2025 Palisades Fire, linking 12,081 CAL FIRE damage inspections to pre-fire Sentinel-2 vegetation amount and moisture, Landsat surface temperature, LANDFIRE fuels, 10 m terrain, and a dated OpenStreetMap representation of buildings and roads. Residential destruction (5,566 of 9,883 inspected residential structures) was modeled with nested logistic and gradient-boosting models evaluated under random and 1 km spatial block cross-validation. Random cross-validation suggested excellent discrimination for the integrated gradient-boosting model (ROC-AUC 0.92), but spatial validation reduced performance to 0.75; an interpretable logistic model performed equivalently under spatial blocking and was better calibrated. Neighborhood building count was the strongest predictor: destruction odds increased fourfold per standard deviation in buildings within 100 m, with a nonlinear increase above approximately 50-60 buildings. Vegetation moisture and amount carried opposing conditional signals: the Normalized Difference Moisture Index at 100-300 m was protective (odds ratio 0.52), whereas the Normalized Difference Vegetation Index at 30-100 m was positively associated with destruction after accounting for moisture (odds ratio 1.74). Predictive information was concentrated at the 100-300 m neighborhood scale. A separate impact track mapped burn severity and 18-month vegetation recovery without introducing post-fire leakage; tract-level social vulnerability showed no detectable association within the study area\textquotesingle s restricted low-vulnerability range. The results position open geospatial modeling as a neighborhood-scale screening tool rather than a parcel-level prediction system. For planning, the actionable signal lies in dense urban fabric, moisture-aware vegetation management, and calibrated uncertainty; for research, spatial block validation should be a minimum standard in single-event wildfire susceptibility studies. The complete workflow is designed for rapid transfer to other damage-inspection datasets and supports reproducible, decision-ready urban resilience assessment.
\end{abstract}


\maketitle

\section{Introduction}\label{introduction}

Wind-driven fires that penetrate dense residential districts are not
simply larger wildfires. Once one or more buildings ignite, structures
become fuel, local ember production intensifies, and fire can propagate
through the built fabric after the initial wildland flame front has
passed \citep{cohen2000,calkin2023,metz2024}. These
events are therefore urban-system failures as well as ecological
disturbances. Their consequences depend on the arrangement and condition
of buildings, vegetation, roads, and services, and on whether an urban
system can resist loss, sustain essential functions, and recover \citep{meerow2016,mcwethy2019}.

Exposure to this hazard is increasing as the wildland-urban interface
expands. In the United States, development has continued to extend into
fire-prone landscapes, while global mapping shows that the interface
between settlements and wildland vegetation is now widespread across
biomes and continents \citep{radeloff2018,carlson2022,schug2023}. Within affected communities, structure loss is
rarely explained by one factor. California studies repeatedly identify
housing arrangement, local structure density, and location as major
correlates of loss, often with effects that rival or exceed coarse
measures of vegetation cover \citep{syphard2012,kramer2019,syphard2019}. The home-ignition zone remains important, but
community-scale morphology also matters because radiant heat, ember
exchange, and repeated ignition opportunities operate across neighboring
parcels \citep{cohen2000,calkin2014,syphard2014,knapp2021,syphard2021}.

Vegetation is an equally necessary but easily oversimplified component
of this system. Green space can moderate heat, stabilize soils, and
support ecological function, yet dry, continuous biomass can increase
combustible exposure. Greenness, canopy cover, vegetation moisture, fuel
type, and spatial arrangement are therefore not interchangeable.
Moisture-sensitive optical indices have been linked to live fuel
moisture in southern California shrublands \citep{dennison2005}, and
recent California analyses show that the relationship between urban
vegetation and structure loss varies with vegetation type, maintenance,
spatial scale, and built context \citep{escobedo2025,kenny2026}. This distinction matters for planning: a greenness-only map can
conflate actively transpiring vegetation with desiccated fuel. At leaf
scale, related spectral work found that water-absorption features can
reveal physiological stress before later-stage canopy interference
becomes dominant \citep{narimani2025b}.

Urban informatics provides a useful frame for integrating these
relationships without reducing them to a stack of disconnected maps. The
field combines urban science, computing, and GIScience to organize
heterogeneous observations into spatially explicit, decision-oriented
systems \citep{batty2013,shi2022}. Open Earth-observation
products, administrative inspections, and volunteered geographic
information can make rapid post-event analysis feasible, but decision
readiness depends on transparent provenance, scale-aware modeling,
reproducible evaluation, and a clear path from diagnostics to
intervention \citep{pradeep2026}. Related work by the authors has emphasized
both the transferability limits of Sentinel-2 and machine-learning
workflows and the value of updateable, neighborhood-scale urban-canopy
screening with explicit spatial diagnostics \citep{narimani2026b,narimani2026d}. Data abundance alone does not guarantee reliable planning
evidence.

The January 2025 Los Angeles fires have already generated important
empirical work. \citet{kenny2026} found structure density to be the
strongest predictor of home loss across the Eaton and Palisades fires,
with smaller and less consistent urban-canopy effects. \citet{norlen2026} likewise showed that urban morphology was more informative than
vegetation cover and that relationships changed across community,
neighborhood, and parcel scales. Those studies established the central
importance of urban form, but they did not combine a temporally
disciplined multi-sensor environmental stack with out-of-area spatial
validation and probability calibration. The Palisades event therefore
still offers an opportunity to ask not only which conditions were
associated with loss, but also how honestly open geospatial models can
estimate transferable susceptibility within a single fire.

This distinction is consequential because spatially autocorrelated
predictors and outcomes can make random train-test splits appear far
more accurate than they are. Nearby observations share landscape,
neighborhood, and fire-exposure histories; placing them on both sides of
a random split allows a model to learn local signatures rather than
transferable relationships \citep{roberts2017,valavi2019,ploton2020}. For planning applications, discrimination alone is
also insufficient. A map that ranks neighborhoods well but assigns
overconfident probabilities can misdirect resources and invite
parcel-level interpretations that the data do not support.

This study asks which pre-fire environmental, topographic, and
built-environment conditions were associated with structure destruction
among CAL FIRE-inspected structures in the Palisades Fire, and how well
those relationships transfer across space. We assemble a fully open
predictor stack with strict temporal separation; compare interpretable
logistic regression and gradient boosting under random and spatially
blocked validation; distinguish vegetation amount from vegetation
moisture across three spatial scales; and maintain a separate
impact-and-context track for burn severity, ecological recovery, social
vulnerability, and service accessibility. We do not collapse these
dimensions into an arbitrary resilience index. Instead, we treat
structure resistance, post-fire ecological response, and community
context as related but non-interchangeable parts of urban wildfire
resilience. The objective is a reproducible neighborhood-screening
framework that makes both actionable signals and unresolved uncertainty
visible.

\section{Materials and methods}\label{materials-and-methods}

\subsection{Study design and urban-informatics architecture}\label{study-design-and-urban-informatics-architecture}

The workflow was organized as a data-to-decision system rather than a
conventional layer-overlay exercise (Figure 1). The primary analytical
track estimates associations with observed structure destruction using
pre-fire information only. A separate track characterizes event impact,
early ecological response, and community context. This separation was
declared before modeling so that burn severity, post-fire spectral
change, observed damage, and recovery could not leak into the
susceptibility predictors. The architecture links six functions of urban
informatics: urban science, heterogeneous urban data, field and remote
sensing, spatial computing, systems interpretation, and a planning
application pathway.

\begin{figure*}[H]
\centering
\includegraphics[width=\textwidth]{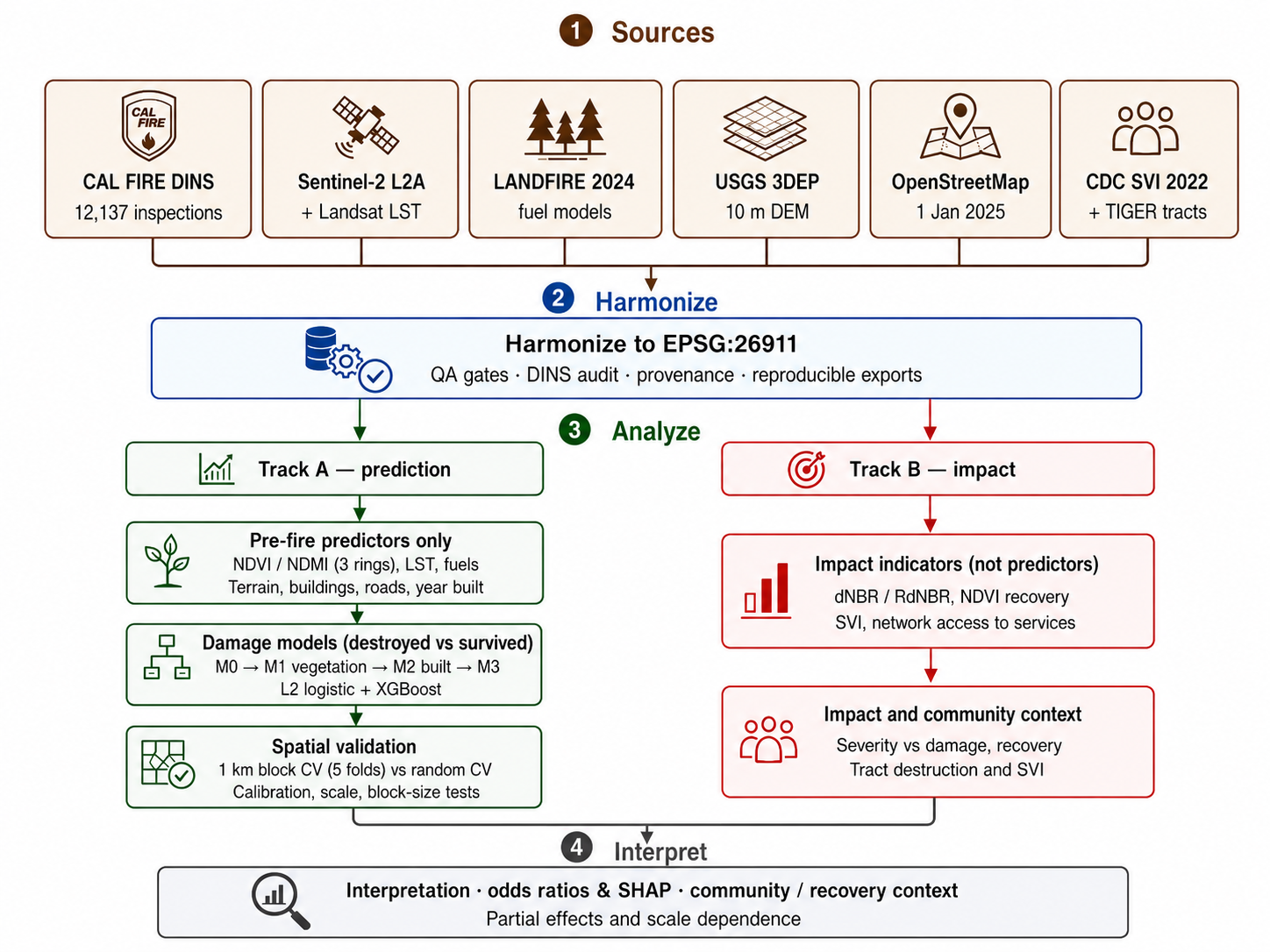}
\caption{Analytical workflow. Public data are harmonized to NAD83 / UTM zone 11N (EPSG:26911), quality controlled, and carried into two complementary analytical tracks. Track A uses only pre-fire predictors to model residential destruction, while Track B interprets event impact and community context without using post-fire indicators as predictors.}
\end{figure*}
\vspace{0.35\baselineskip}

The individual CAL FIRE-inspected structure is the primary analysis
unit. All metric operations were performed in NAD83 / UTM zone 11N
(EPSG:26911). Three spatial supports were used for environmental and
neighborhood variables: 0-30 m, 30-100 m, and 100-300 m. These supports
represent an immediate structure-adjacent zone, an intermediate
neighborhood zone, and a broader community context. Because 10-30 m
satellite pixels cannot resolve individual shrubs, fences, or
yard-maintenance details, the innermost support is interpreted as a
local remote-sensing context, not as a parcel-level defensible-space
inspection.

\subsection{Study area and event context}\label{study-area-and-event-context}

The Palisades Fire ignited on 7 January 2025 in the Santa Monica
Mountains above Pacific Palisades, Los Angeles, California, and burned
approximately 23,448 acres (about 95 km2) before containment at the end
of January \citep{calfire2025incident}. The study area comprises the final
interagency fire perimeter plus a 2 km buffer, covering approximately
218 km2 \citep{nifc2025wfigs} (Figure 2). The buffer
retains inspected structures near the mapped perimeter and provides
environmental and accessibility context without treating uninspected
buildings as undamaged.

\begin{figure*}[H]
\centering
\includegraphics[width=\textwidth]{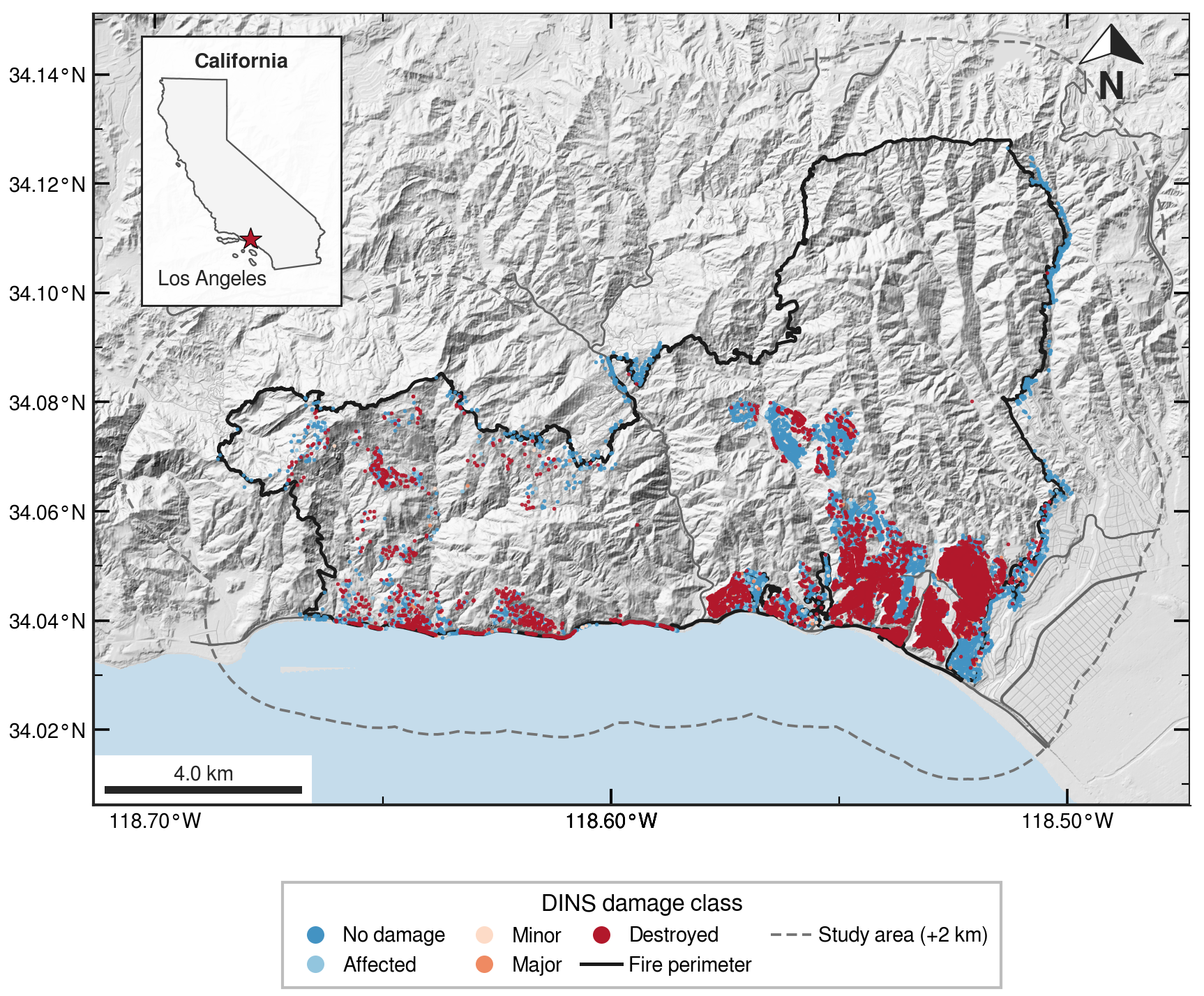}
\caption{Study area and observed structure damage. Final Palisades Fire perimeter, 2-km study-area buffer, terrain and road context, and CAL FIRE DINS-inspected residential structures colored by original damage class. The inset locates the fire in California.}
\end{figure*}
\vspace{0.35\baselineskip}

The event followed exceptional antecedent dryness and coincided with
strong Santa Ana conditions. Study-area gridMET precipitation for
October-December 2024 totaled 5.5 mm, compared with a 1980-2024 median
of 111 mm. One-hundred-hour dead fuel moisture reached its lowest values
of the preceding year while vapor-pressure deficit remained elevated.
ERA5-Land area means for 7-9 January reached 7.8 m s-1 for 10 m wind
speed and 9.6\% for minimum relative humidity \citep{munozsabater2021}. These coarse meteorological data describe the common forcing
under which structures were exposed; they were not used as
structure-level predictors because their 3-9 km grids provide little
credible spatial contrast across the fire footprint. The event-context
series is shown in Figure 3. The event is interpreted as a
wind-dominated fire transitioning from chaparral into dense urban fabric
\citep{guzman2019,keeley2019}.

\begin{figure*}[H]
\centering
\includegraphics[width=\textwidth]{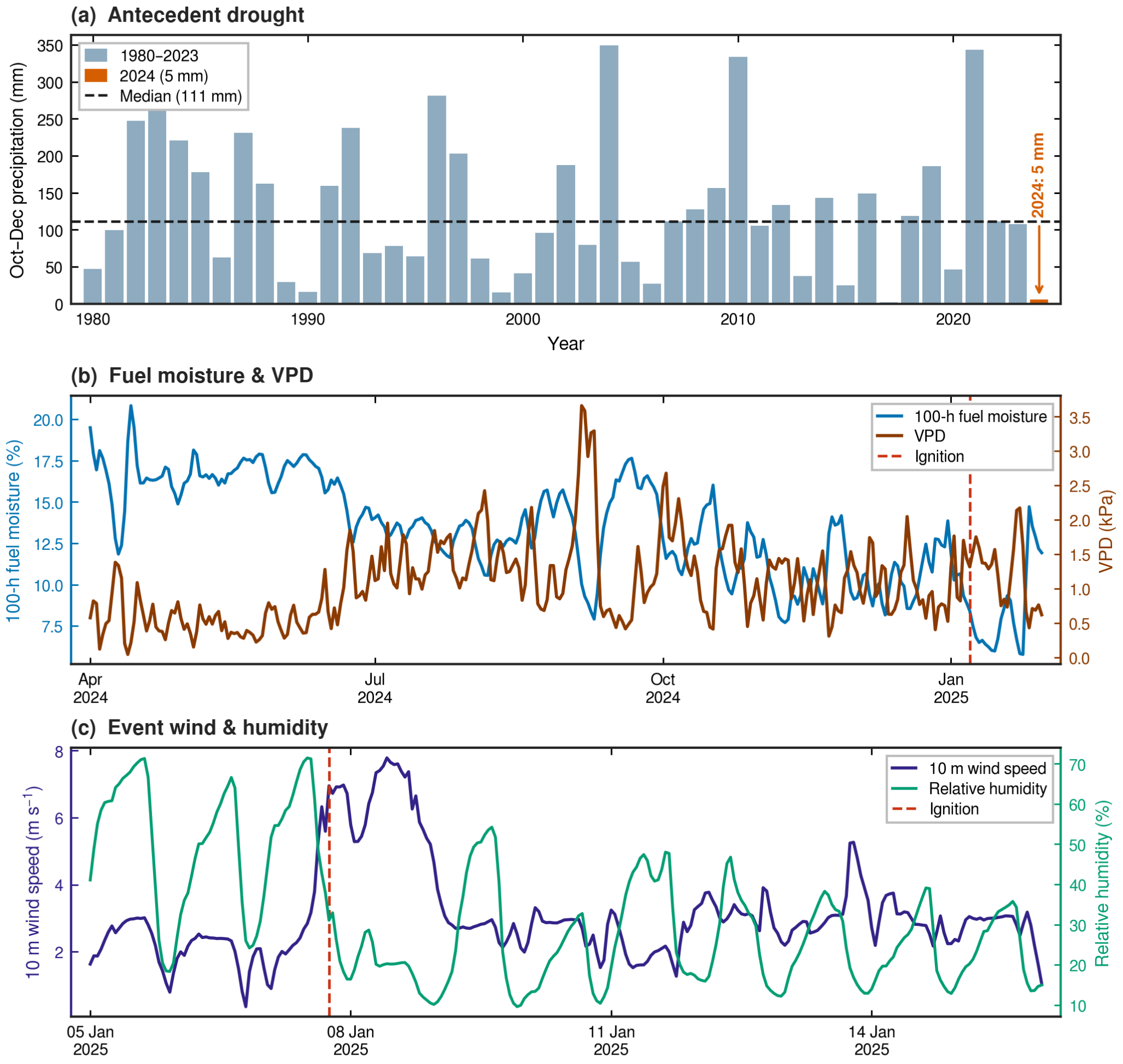}
\caption{Antecedent and event fire-weather context. (A) October-December precipitation totals for 1980-2024 with the historical median and the exceptionally dry 2024 season highlighted; (B) 100-hour dead-fuel moisture and vapor-pressure deficit from April 2024 through January 2025; and (C) 10-m wind speed and relative humidity during the January 2025 event window. These variables summarize the broader climatic and event setting and were not used as structure-level predictors.}
\end{figure*}
\vspace{0.35\baselineskip}

\subsection{Damage observations and analysis population}\label{damage-observations-and-analysis-population}

The primary outcome derives from the CAL FIRE Damage Inspection program,
which classifies inspected structures as No Damage, Affected
(\textgreater0-10\%), Minor (10-25\%), Major (25-50\%), Destroyed
(\textgreater50\%), or Inaccessible \citep{calfire2025dins}. The authoritative
Palisades incident service contained 12,137 records when retrieved on 12
August 2026. Fifty-six inaccessible records were excluded, leaving
12,081 classified structures. Residential structures - Single Residence,
Multiple Residence, and Mixed Commercial/Residential - formed the
primary modeling population (n = 9,883), of which 5,566 (56.3\%) were
destroyed. The binary outcome was therefore Destroyed versus Not
Destroyed (No Damage through Major). A predeclared sensitivity analysis
grouped Major with Destroyed. Intermediate damage classes were retained
for descriptive figures but were too sparse for stable ordinal modeling
(Table 1B).

Inspections were treated as the sampling frame. Structures outside the
inspected set were considered unknown, never undamaged. DINS also
records roof, eave, vent-screen, siding, and window characteristics
observed after the fire. Those fields were excluded because
observability depended on the outcome: for example, eave type was
unknown for 69.2\% of destroyed structures and 0.3\% of undamaged
structures. Including such variables would allow the model to infer
destruction from the absence of inspectable material, a form of outcome
leakage also noted in post-fire engineering assessments \citep{ibhs2025}. Assessor-derived year
built and improved value were retained because they existed
independently of the post-fire inspection and were 100\% and 96.5\%
complete, respectively.

\begin{table*}[t]
\centering
\textbf{Table 1.} Public data architecture and audited analysis
population. Section A reproduces the complete data-source inventory;
Section B reproduces the full DINS damage-by-structure-category audit.
\scriptsize
\setlength{\tabcolsep}{3pt}
\textbf{A. Public data sources}
\vspace{0.15\baselineskip}
\begin{tabular*}{\textwidth}{@{\extracolsep{\fill}}>{\raggedright\arraybackslash}p{0.137\textwidth}>{\raggedright\arraybackslash}p{0.137\textwidth}>{\raggedright\arraybackslash}p{0.095\textwidth}>{\raggedright\arraybackslash}p{0.208\textwidth}>{\raggedright\arraybackslash}p{0.194\textwidth}>{\raggedright\arraybackslash}p{0.204\textwidth}@{}}
\toprule

\textbf{Dataset}
 & \textbf{Provider}
 & \textbf{Resolution}
 & \textbf{Temporal window used}
 & \textbf{Variables derived}
 & \textbf{Analytical role}
 \\
\midrule
CAL FIRE DINS & CAL FIRE / California Open Data & Point (structure) &
Post-fire inspections, 2025 & Damage class, structure category, year
built & Primary outcome \\
WFIGS Interagency Fire Perimeters & NIFC & Polygon & Final 2025
perimeter & Fire perimeter geometry & Study area definition \\
Sentinel-2 L2A (via GEE) & Copernicus/ESA & 10-20 m & Oct 2024-Jan 2025
(pre); Oct-Dec 2022-2024 (baseline); Jan-F & NDVI, NDMI, NBR/dNBR &
Pre-fire vegetation predictors; severity \& recovery \\
Landsat 8/9 C2 L2 (via GEE) & USGS & 30 m & Oct 2024-Jan 2025 & Land
surface temperature & Pre-fire thermal predictor \\
LANDFIRE LF2024 & USGS/USDA-FS & 30 m & 2024 update (pre-fire vintage) &
FBFM40 fuel models, canopy cover & Fuel type/continuity predictors \\
USGS 3DEP DEM (via GEE) & USGS & 10 m & Static & Elevation, slope,
aspect, TPI, TRI & Terrain predictors \\
OpenStreetMap (Overpass attic) & OSM contributors (ODbL) & Feature &
Snapshot 2025-01-01 (pre-fire) & Buildings, roads, facilities &
Built-environment predictors; accessibility \\
gridMET (via GEE) & Climatology Lab & \textasciitilde4 km daily & Apr
2024-Jan 2025; Oct-Dec 1980-2024 & Precipitation, VPD, fuel moisture,
ERC & Antecedent climate context \\
ERA5-Land (via GEE) & ECMWF/Copernicus & \textasciitilde9 km hourly &
Jan 5-15, 2025 & Wind, temperature, dewpoint & Event weather context \\
CDC/ATSDR SVI 2022 & CDC/ATSDR & Census tract & 2022 release &
Vulnerability themes, vehicle access, age & Community context (Track B) \\
TIGER/Line 2023 & U.S. Census Bureau & Tract/block group & 2023 vintage
& Census geometries & Community units \\
\bottomrule
\end{tabular*}
\vspace{0.35\baselineskip}
\textbf{B. CAL FIRE DINS audit by damage class and structure category}
\vspace{0.15\baselineskip}
\begin{tabular*}{\textwidth}{@{\extracolsep{\fill}}>{\raggedright\arraybackslash}p{0.129\textwidth}>{\raggedright\arraybackslash}p{0.083\textwidth}>{\raggedright\arraybackslash}p{0.148\textwidth}>{\raggedright\arraybackslash}p{0.106\textwidth}>{\raggedright\arraybackslash}p{0.152\textwidth}>{\raggedright\arraybackslash}p{0.134\textwidth}>{\raggedright\arraybackslash}p{0.125\textwidth}>{\raggedright\arraybackslash}p{0.074\textwidth}@{}}
\toprule

\textbf{DAMAGE}
 & \textbf{Infrastructure}
 & \textbf{Mixed Commercial/Residential}
 & \textbf{Multiple Residence}
 & \textbf{Nonresidential Commercial}
 & \textbf{Other Minor Structure}
 & \textbf{Single Residence}
 & \textbf{All}
 \\
\midrule
Affected (\textgreater0-10\%) & 5 & 1 & 37 & 27 & 81 & 581 & 732 \\
Destroyed (\textgreater50\%) & 1 & 3 & 135 & 158 & 1120 & 5428 & 6845 \\
Inaccessible & 0 & 0 & 0 & 0 & 6 & 50 & 56 \\
Major (25-50\%) & 0 & 0 & 7 & 5 & 9 & 51 & 72 \\
Minor (10-25\%) & 0 & 1 & 8 & 7 & 21 & 134 & 171 \\
No Damage & 56 & 6 & 228 & 146 & 562 & 3263 & 4261 \\
All & 62 & 11 & 415 & 343 & 1799 & 9507 & 12137 \\
\bottomrule
\end{tabular*}
\end{table*}
\subsection{Open data architecture and feature engineering}\label{open-data-architecture-and-feature-engineering}

Table 1A summarizes the public data architecture, Table 1B audits the
DINS analysis population, and Table 2 lists the complete predictor
dictionary. Three temporal windows were kept distinct: an immediate
pre-fire period (1 October 2024-6 January 2025), a season-matched
baseline (October-December 2022-2024), and post-fire observations used
only in the impact track. The use of temporally consistent composites
follows a broader remote-sensing lesson: rich spectral inputs do not
ensure transferability when acquisition windows, ground reference, and
spatial validation are poorly aligned \citep{narimani2026d}. A
related Sentinel-2 time-series workflow likewise used phenological
alignment and moisture-sensitive indices to preserve stress signals
across growth stages \citep{narimani2025a}. True- and
shortwave-infrared false-color scenes provide pre-fire, active-fire, and
post-fire visual context (Figure 4), while pre-fire predictor maps and
post-fire impact maps are presented separately in Figures 5 and 6,
respectively.

\begin{figure*}[H]
\centering
\includegraphics[width=\textwidth]{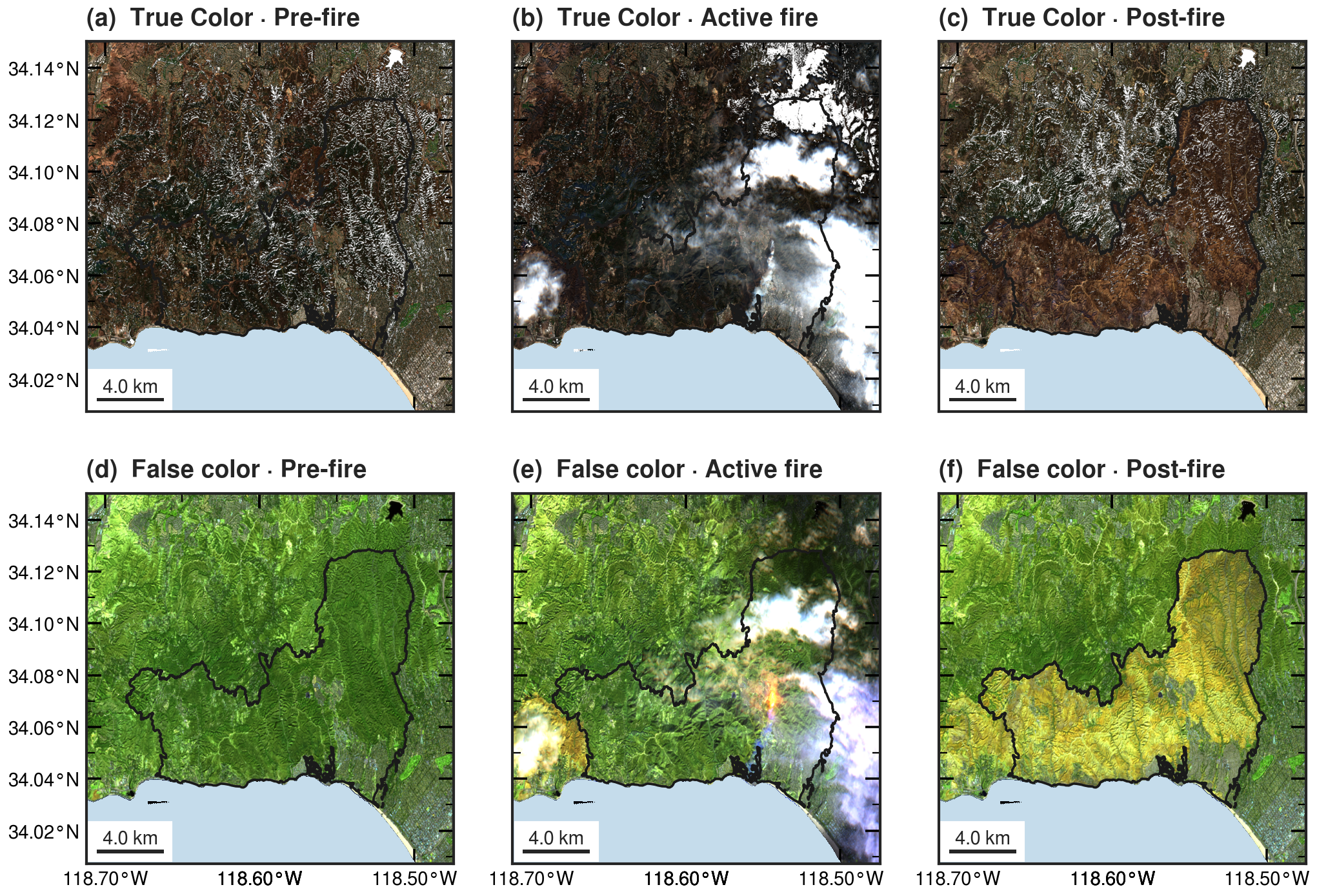}
\caption{Sentinel-2 visual context before, during, and after the fire. The top row shows true-color composites and the bottom row shows shortwave-infrared false-color composites for pre-fire, active-fire, and post-fire conditions. The active-fire panels retain smoke and shortwave-infrared hotspots.}
\end{figure*}
\vspace{0.35\baselineskip}

\begin{figure*}[H]
\centering
\includegraphics[width=\textwidth]{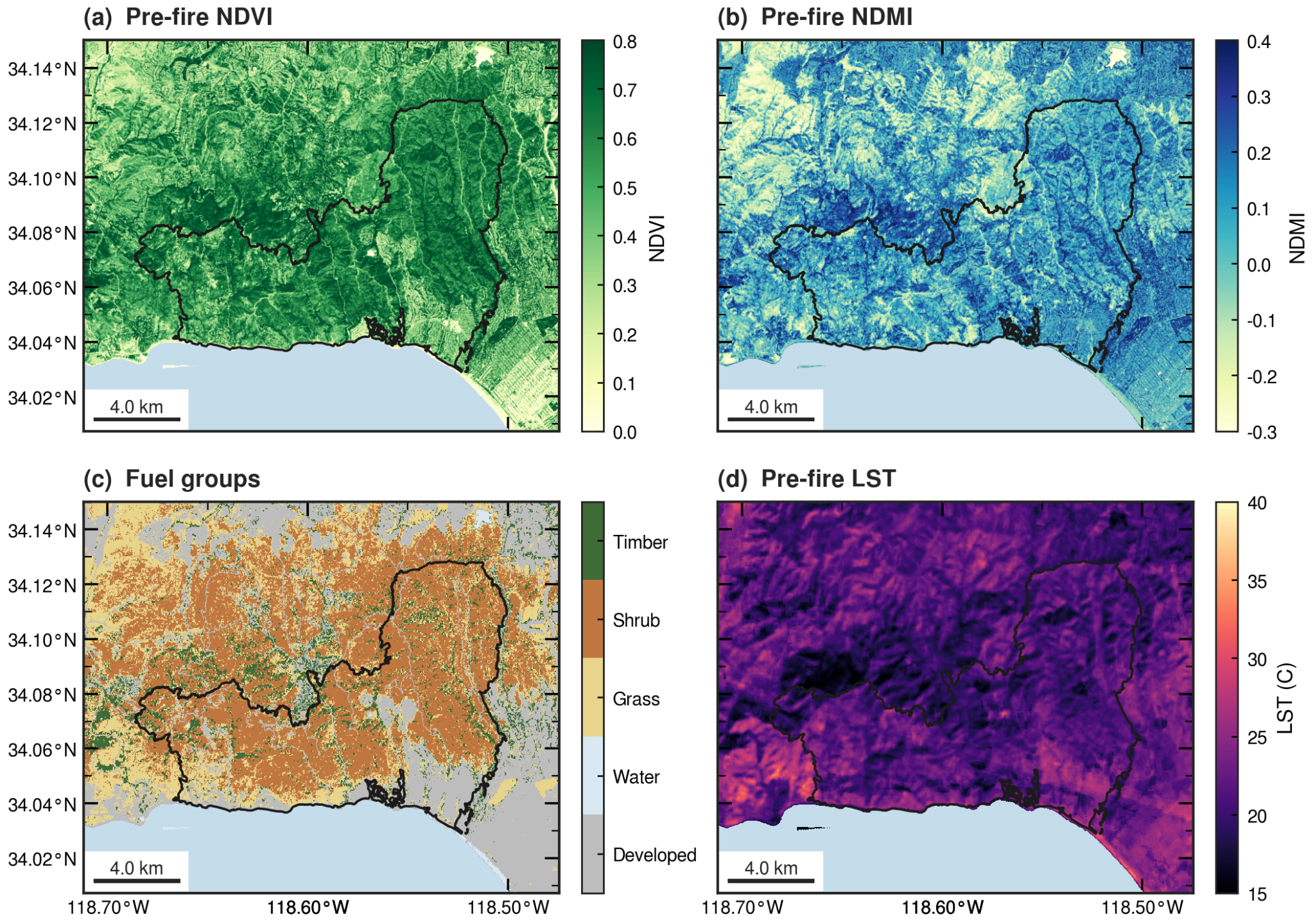}
\caption{Pre-fire environmental conditions. (A) Pre-fire NDVI; (B) pre-fire NDMI; (C) LANDFIRE fuel groups; and (D) Landsat land-surface temperature. These variables summarize the pre-fire vegetation and thermal setting used in the predictive track.}
\end{figure*}
\vspace{0.35\baselineskip}

\begin{figure*}[H]
\centering
\includegraphics[width=\textwidth]{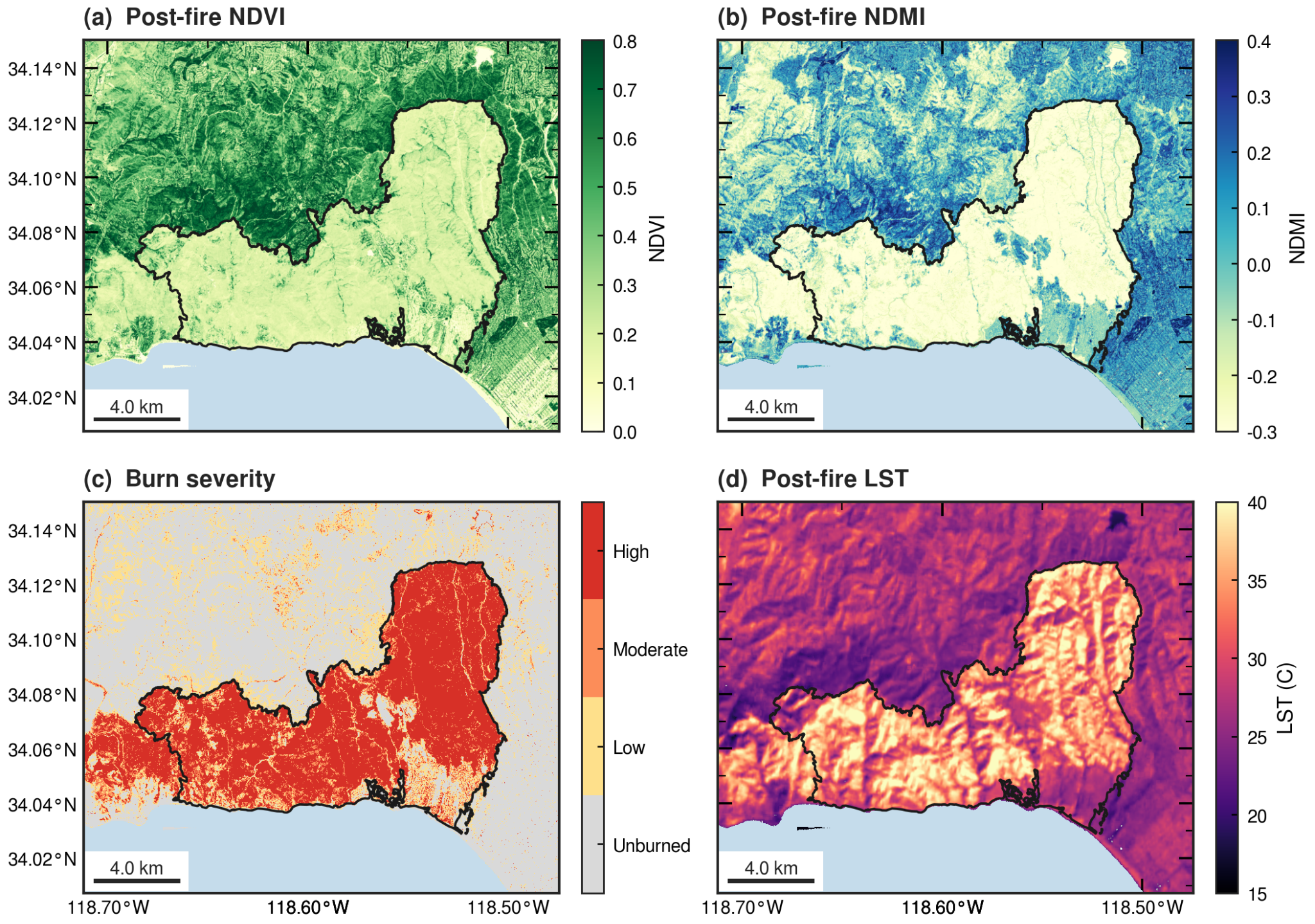}
\caption{Post-fire environmental conditions. (A) Post-fire NDVI; (B) post-fire NDMI; (C) study-derived burn-severity classes; and (D) post-fire land-surface temperature. These are impact indicators used only in the post-fire interpretation track.}
\end{figure*}
\vspace{0.35\baselineskip}

\begin{table*}[t]
\centering
\textbf{Table 2.} Complete predictor dictionary, source, and spatial
support.
\footnotesize
\setlength{\tabcolsep}{3pt}
\begin{tabular*}{\textwidth}{@{\extracolsep{\fill}}>{\raggedright\arraybackslash}p{0.198\textwidth}>{\raggedright\arraybackslash}p{0.198\textwidth}>{\raggedright\arraybackslash}p{0.198\textwidth}>{\raggedright\arraybackslash}p{0.198\textwidth}>{\raggedright\arraybackslash}p{0.198\textwidth}@{}}
\toprule

\textbf{Block}
 & \textbf{Variable}
 & \textbf{Definition}
 & \textbf{Source}
 & \textbf{Support}
 \\
\midrule
Terrain (M0) & elevation\_pt & Elevation at structure (m) & 3DEP 10 m & point \\
 & slope\_deg\_pt & Slope (degrees) & 3DEP 10 m & point \\
 & northness\_pt & cos(aspect) & 3DEP 10 m & point \\
 & eastness\_pt & sin(aspect) & 3DEP 10 m & point \\
 & tpi300\_pt & Topographic position index (300 m) & 3DEP 10
m & point \\
 & tri\_pt & Terrain ruggedness index & 3DEP 10 m & point \\
 & dist\_wildland\_m & Distance to nearest wildland fuel
pixel (m) & LANDFIRE FBFM40 & point \\
Vegetation \& fuels (M1) & ndvi\_pre\_r0\_30 & NDVI & Sentinel-2 10 m & 0-30 m ring \\
 & ndvi\_pre\_r30\_100 & NDVI & Sentinel-2 10 m & 30-100 m ring \\
 & ndvi\_pre\_r100\_300 & NDVI & Sentinel-2 10 m & 100-300 m ring \\
 & ndmi\_pre\_r0\_30 & NDMI & Sentinel-2 10 m & 0-30 m ring \\
 & ndmi\_pre\_r30\_100 & NDMI & Sentinel-2 10 m & 30-100 m ring \\
 & ndmi\_pre\_r100\_300 & NDMI & Sentinel-2 10 m & 100-300 m ring \\
 & ndvi\_anom\_r0\_30 & NDVI anomaly (pre minus 2022-2024 baseline) & Sentinel-2 10 m & 0-30 m ring \\
 & ndvi\_anom\_r30\_100 & NDVI anomaly (pre minus 2022-2024 baseline) & Sentinel-2 10 m & 30-100 m ring \\
 & ndvi\_anom\_r100\_300 & NDVI anomaly (pre minus 2022-2024 baseline) & Sentinel-2 10 m & 100-300 m ring \\
 & ndmi\_anom\_r0\_30 & NDMI anomaly (pre minus 2022-2024 baseline) & Sentinel-2 10 m & 0-30 m ring \\
 & ndmi\_anom\_r30\_100 & NDMI anomaly (pre minus 2022-2024 baseline) & Sentinel-2 10 m & 30-100 m ring \\
 & ndmi\_anom\_r100\_300 & NDMI anomaly (pre minus 2022-2024 baseline) & Sentinel-2 10 m & 100-300 m ring \\
 & lst\_pre\_r30\_100 & Land surface temperature
(deg C) & Landsat 30 m & 30-100 m ring \\
 & lst\_pre\_r100\_300 & Land surface
temperature (deg C) & Landsat 30 m & 100-300 m ring \\
 & canopy\_cover\_r0\_30 & Canopy cover (\%) & LANDFIRE 30 m & 0-30 m ring \\
 & canopy\_cover\_r30\_100 & Canopy cover (\%) & LANDFIRE 30 m & 30-100 m ring \\
 & canopy\_cover\_r100\_300 & Canopy cover (\%) & LANDFIRE 30 m & 100-300 m ring \\
 & fuel\_wildland\_r0\_30 & Wildland burnable
fuel fraction & LANDFIRE 30 m & 0-30 m ring \\
 & fuel\_wildland\_r30\_100 & Wildland burnable
fuel fraction & LANDFIRE 30 m & 30-100 m ring \\
 & fuel\_wildland\_r100\_300 & Wildland burnable
fuel fraction & LANDFIRE 30 m & 100-300 m ring \\
Built environment (M2) & nn\_building\_dist\_m & Edge distance to
nearest neighboring building (m) & OSM buildings 2025-01-01 & structure \\
 & footprint\_area\_m2 & Matched building
footprint area (m2) & OSM buildings & structure \\
 & bld\_count\_r30 & Building count within 30 m & OSM buildings & 30 m disk \\
 & bld\_count\_r100 & Building count within 100 m & OSM buildings & 100 m disk \\
 & bld\_count\_r300 & Building count within 300 m & OSM buildings & 300 m disk \\
 & dist\_road\_m & Distance to nearest drivable
road (m) & OSM roads 2025-01-01 & point \\
 & road\_len\_r300\_m\_per\_km2 & Road length
density within 300 m & OSM roads & 300 m disk \\
 & dist\_deadend\_m & Distance to nearest dead-end
node (m) & OSM roads & point \\
 & yearbuilt & Year built (assessor) & DINS/assessor & structure \\
 & assessed\_value\_log & log10 assessed improved
value & DINS/assessor & structure \\
\bottomrule
\end{tabular*}
\end{table*}
Sentinel-2 Level-2A surface reflectance was processed in Google Earth
Engine \citep{gorelick2017} using s2cloudless probability below 40\%
and scene-classification masking. The immediate pre-fire composite used
16 scenes and the seasonal baseline used 42 scenes. The Normalized
Difference Vegetation Index (NDVI; Tucker, 1979) and Normalized
Difference Moisture Index (NDMI; Gao, 1996) were calculated using Eqs.
(1) and (2):

\begin{equation}
NDVI = \frac{\rho_{NIR} - \rho_{Red}}{\rho_{NIR} + \rho_{Red}}
\end{equation}

\begin{equation}
NDMI = \frac{\rho_{NIR} - \rho_{SWIR}}{\rho_{NIR} + \rho_{SWIR}}
\end{equation}

where $\rho_{NIR}$, $\rho_{Red}$, and $\rho_{SWIR}$ denote surface reflectance in the
near-infrared, red, and shortwave-infrared bands, respectively. NDVI
characterizes vegetation greenness, whereas NDMI was used as a
moisture-sensitive indicator of canopy condition rather than as a direct
measurement of fuel moisture \citep{dennison2005}. Index anomalies
were calculated as the immediate pre-fire value minus the season-matched
2022-2024 baseline. Retaining both greenness and moisture indicators is
also consistent with multi-trait and hybrid spectral studies showing
that structural, water-related, and biochemical information can provide
complementary predictive signals \citep{farajpoor2025c,chakraborty2025}.

Landsat 8/9 Collection 2 Level-2 surface temperature, quality masked and
composited across 21 scenes, provided pre-fire land-surface temperature.
LANDFIRE 2024 supplied Scott and Burgan fire-behavior fuel models and
forest canopy cover at 30 m \citep{scott2005,rollins2009,landfire2024}. The 2024 release was selected because later products
may encode the fire scar. Elevation, slope, northness, eastness, 300 m
topographic position, and terrain ruggedness were derived from the USGS
3DEP 10 m digital elevation model.

Built-environment variables came from a date-scoped OpenStreetMap
snapshot for 1 January 2025, six days before ignition \citep{osm2025}. The extract contained 28,302 building polygons and
4,904 drivable road edges; 99.1\% of residential DINS points matched a
building footprint within 20 m. Derived variables included building
footprint area, edge-to-edge distance to the nearest neighboring
building, building counts within 30, 100, and 300 m, distance to the
nearest drivable road and dead-end node, and road-length density within
300 m. Fire-station and hospital locations supported the contextual
accessibility analysis. Census-tract geometry came from TIGER/Line and
social context from the 2022 CDC/ATSDR Social Vulnerability Index
\citep{flanagan2011,cdcsvi2022}.

Raster predictors were summarized as valid-pixel means over concentric
rings of 0-30, 30-100, and 100-300 m. Fuel-model classes were collapsed
into interpretable group fractions, and Euclidean distance to the
nearest wildland-fuel pixel was calculated from LANDFIRE. Collinearity
was assessed with Spearman correlations and variance-inflation factors.
Perfectly dependent fuel subfractions were removed, and the
interpretable model used a pruned predictor set in which pairs with
absolute correlation above 0.85 were resolved using process relevance
and data support. The final integrated set contained 37 predictors
organized into nested blocks: M0, terrain and distance to wildland fuel;
M1, M0 plus vegetation, moisture, thermal, and fuel variables; M2, M0
plus built-environment variables; and M3, all predictor families. This
multi-predictor design is consistent with proximal multi-trait retrieval
work in which shared spectral structure across correlated outcomes was
retained rather than collapsed into a single index \citep{farajpoor2024}.

\subsection{Models, spatial validation, and interpretation}\label{models-spatial-validation-and-interpretation}

Two model families were used to separate interpretability from nonlinear
flexibility. Logistic regression modeled the probability that structure
i was destroyed as shown in Eq. (3):

\begin{equation}
logit\left[ P(Y_{i} = 1)\right] = \beta_{0} + \sum_{j = 1}^{p}{\beta_{j} x_{ij}}
\end{equation}

where $P(Y_i = 1)$ is the probability that structure i was destroyed, $\beta_0$ is
the intercept, $\beta_j$ is the coefficient for standardized predictor $x_{ij}$, and
p is the number of predictors. The logistic model used L2 regularization
and balanced class weights. Standardized odds ratios and model-based
95\% confidence intervals were obtained from an unpenalized refit of the
pruned specification and are interpreted as conditional associations
rather than causal effects.

Gradient-boosted trees were fitted with XGBoost and tuned over tree
depth, learning rate, and number of estimators \citep{chen2016}. Predictor contributions were summarized with SHAP values, which
are interpreted as model attributions rather than causal effects
\citep{lundberg2017}.

The primary evaluation used five-fold spatial block cross-validation.
Structures within the same 1 km grid cell were assigned to the same
fold, reducing direct neighborhood leakage while preserving enough
spatial units for stable evaluation. Observations near adjacent block
edges can remain close, so 500 m and 2 km block sizes were evaluated as
sensitivity bounds. Median imputation, scaling, class weighting, and
inner three-fold hyperparameter tuning were performed within each
training fold. Conventional random five-fold cross-validation was
repeated only to quantify optimism from ignoring spatial dependence
\citep{roberts2017,valavi2019,ploton2020}. The
emphasis on domain-aware evaluation is also consistent with work showing
that instrument and domain shifts can propagate error through hybrid
spectral models \citep{farajpoor2025a}.

Performance was evaluated with ROC-AUC, precision-recall AUC, balanced
accuracy, F1 score, and Brier score; precision-recall AUC was retained
because the classes were moderately imbalanced \citep{saito2015}. Reliability curves and logistic recalibration intercepts and
slopes assessed probability calibration. Fold-level means and standard
deviations were reported, and Moran' s I was calculated
for out-of-fold residuals. Sensitivity analyses varied ring support,
spatial block size, outcome definition, and population. A model-response
inflection near a given density is treated as a screening signal, not as
a causal or regulatory threshold.

\subsection{Impact, recovery, and community context}\label{impact-recovery-and-community-context}

The impact track was analytically separated from the pre-fire models.
Burn severity was derived from pre- and post-fire Sentinel-2 Normalized
Burn Ratio (NBR) composites. NBR, differenced NBR (dNBR), and relative
dNBR (RdNBR) were calculated using Eqs. (4)-(6) \citep{key2006,miller2007}:

\begin{equation}
NBR = \frac{\rho_{NIR} - \rho_{SWIR2}}{\rho_{NIR} + \rho_{SWIR2}}
\end{equation}

\begin{equation}
dNBR = NBR_{pre} - NBR_{post}
\end{equation}

\begin{equation}
RdNBR = \frac{dNBR}{\left( \frac{\left| NBR_{pre} \right|}{1000} \right)^{\frac{1}{2}}}
\end{equation}

where $\rho_{NIR}$ and $\rho_{SWIR2}$ are surface reflectance in the near-infrared and
longer-wave shortwave-infrared bands; $NBR_{pre}$ and $NBR_{post}$ are the pre-
and post-fire NBR composites; $dNBR$ is their difference; and $RdNBR$ scales
$dNBR$ by the magnitude of pre-fire NBR to reduce dependence on initial
vegetation condition.

Conventional thresholds of 100, 270, and 440 on the scaled dNBR were
used to label low, moderate, and high severity. Because urban spectral
change combines vegetation loss, structural change, ash, and exposed
soil, severity is presented as an impact descriptor rather than a direct
measure of structure-level fire intensity.

Vegetation response was tracked monthly from February 2025 through July
2026. To separate post-fire change from seasonal phenology, relative
NDVI recovery for month m and burn-severity class s was calculated using
Eq. (7):

\begin{equation}
R_{m,s} = \frac{mean(NDVI_{2025,m,s})}{mean(NDVI_{2022 - 2024,m,s})}
\end{equation}

where $R_{m,s}$ is the relative NDVI recovery ratio; $\mathrm{mean}(NDVI_{2025,m,s})$ is
the mean post-fire NDVI for month m and severity class s; and
$\mathrm{mean}(NDVI_{2022-2024,m,s})$ is the month-matched 2022-2024 climatological
mean for the same pixels. A value of 1 indicates recovery to the
seasonal baseline.

Community context joined tract-level residential destruction fractions
to the CDC/ATSDR Social Vulnerability Index and to median pre-fire
network distance from inspected structures to the nearest fire station.
The accessibility measure describes baseline service proximity rather
than evacuation performance or emergency response time. The tract
analysis is descriptive because only twelve tracts contained inspected
structures.

\subsection{Reproducibility and quality assurance}\label{reproducibility-and-quality-assurance}

The workflow was implemented in Python using geopandas, rasterio,
scikit-learn, statsmodels, XGBoost, SHAP, osmnx, libpysal, and esda,
with Google Earth Engine used for satellite compositing. Every source
was recorded in a manifest with provider, version, spatial and temporal
support, access parameters, retrieval time, license, and checksum.
Deterministic seeds fixed fold assignments and model tuning. Automated
assertions tested coordinate reference systems, unique structure
identifiers, expected index ranges, temporal precedence, and the
exclusion of post-fire or observability-biased variables from the
predictor matrix. Analysis-ready data and derived products are archived
in Zenodo \citep{palisades2026data}, and the public replication
code is available at
\url{https://github.com/MohammadrezaNarimaniUCDavis/Palisades_Urban_Wildfire_GeoAI}.

\section{Results}\label{results}

\subsection{Event setting and spatial pattern of loss}\label{event-setting-and-spatial-pattern-of-loss}

The fire occurred after the second-driest October-December period in the
45-year study-area record and during a sharp combination of low relative
humidity, low dead-fuel moisture, and strong winds (Figure 3). Within
the inspected residential population, 5,566 of 9,883 structures were
destroyed. Loss was strongly clustered (Moran' s I = 0.59) and concentrated in the contiguous urban fabric of Pacific
Palisades and along the Highway 1 corridor. Structures in the northern
and western canyon-edge subdivisions and many inspected structures near
the buffered perimeter experienced markedly lower destruction rates
(Figure 2).

\subsection{Pre-fire environmental and built-environment contrasts}\label{pre-fire-environmental-and-built-environment-contrasts}

The pre-fire environmental maps show a shrub-dominated landscape with
substantial local variation in vegetation moisture and thermal condition
(Figure 5), while the post-fire maps show pronounced changes in
greenness, moisture, burn severity, and surface temperature (Figure 6).
The built-environment maps show the highest structure densities in the
southeastern urban fabric, where many losses occurred (Figure 7). Across
original DINS classes, destroyed structures occupied denser
neighborhoods, were surrounded by lower NDMI, were closer to dead-end
road nodes, occurred at higher local topographic positions, and tended
to be older than structures with no recorded damage (Figure 8).
Nearest-neighbor building distances were short throughout the developed
area (median 2.6 m), so neighborhood building count separated the
classes more clearly than nearest-neighbor distance alone.

\begin{figure*}[H]
\centering
\includegraphics[width=\textwidth]{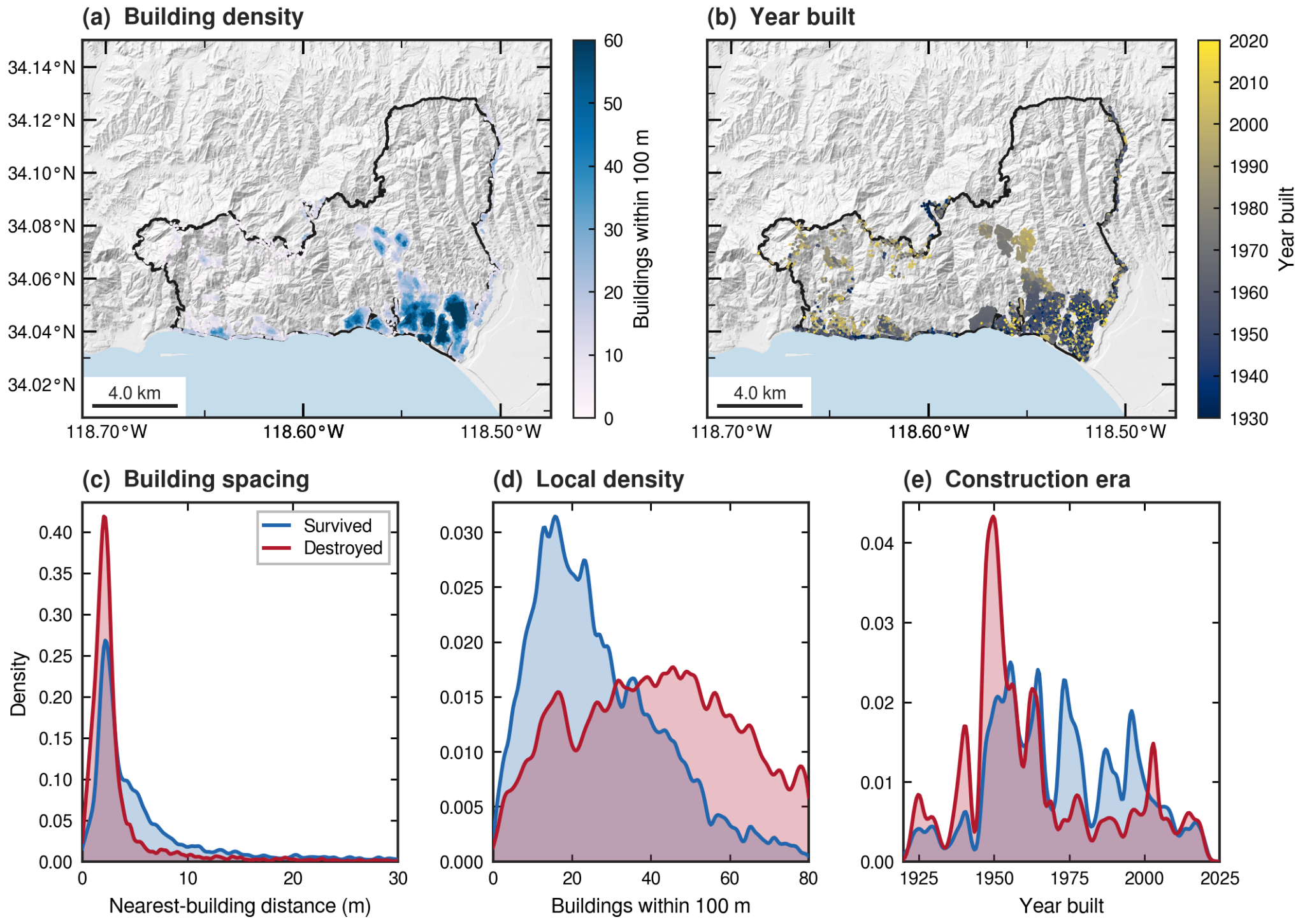}
\caption{Built environment and structure arrangement. (A) Building count within 100 m; (B) assessor year built; and (C-E) distributions comparing nearest-building distance, local building density within 100 m, and construction era for destroyed versus not-destroyed residential structures.}
\end{figure*}
\vspace{0.35\baselineskip}

\begin{figure*}[H]
\centering
\includegraphics[width=\textwidth]{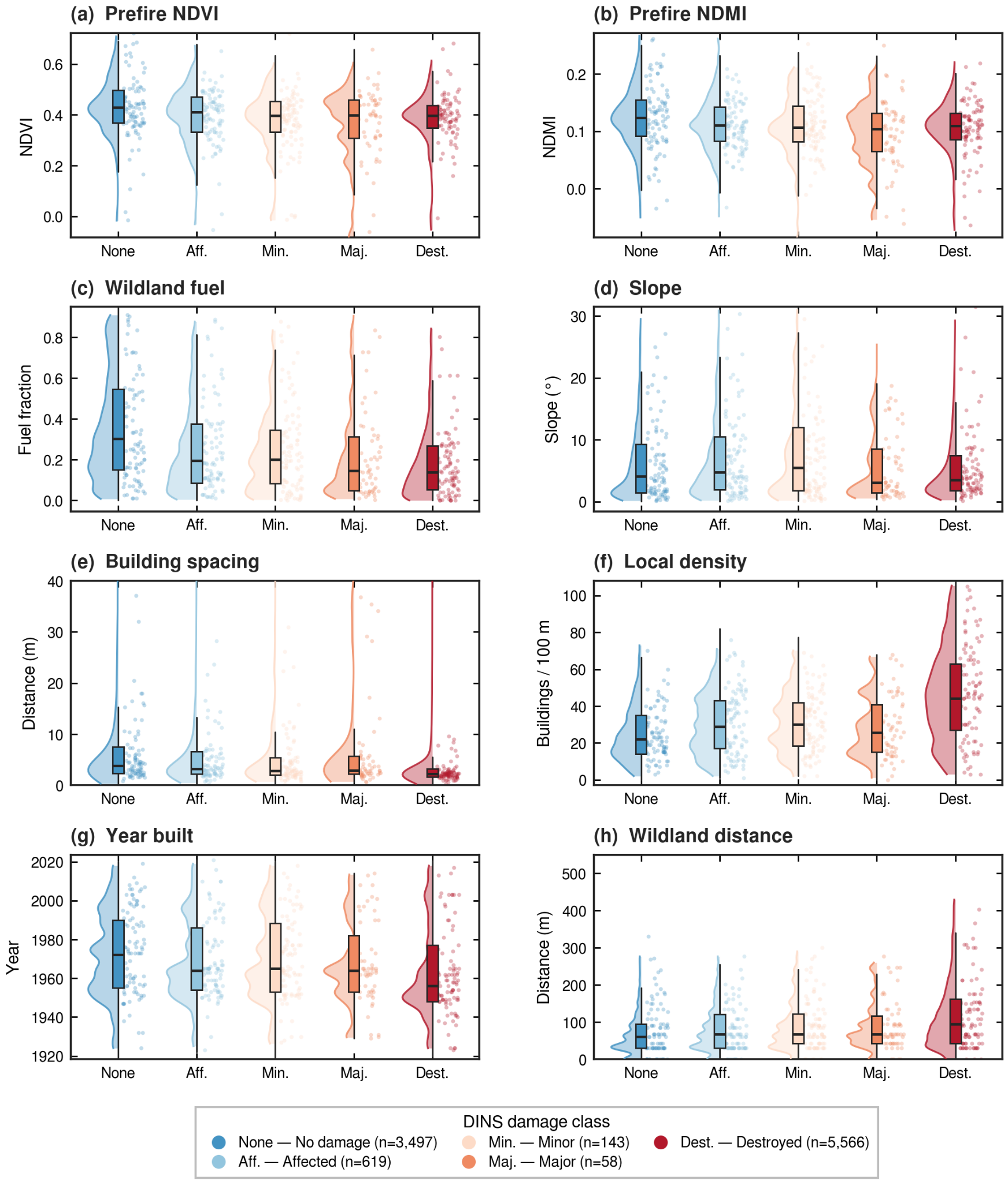}
\caption{Pre-fire predictor distributions by original DINS damage class. Violin-box-scatter panels summarize pre-fire NDVI, pre-fire NDMI, wildland-fuel fraction, slope, nearest-building distance, local building density within 100 m, year built, and distance to wildland fuels across the five DINS damage classes.}
\end{figure*}
\vspace{0.35\baselineskip}

\subsection{Transferable model performance and calibration}\label{transferable-model-performance-and-calibration}

Validation design changed the apparent quality of the models more than
the choice of learner. Under random cross-validation, the integrated
XGBoost model achieved ROC-AUC 0.921 +/- 0.007, precision-recall AUC
0.938, and Brier score 0.112. Under 1 km spatial blocking, the same
specification fell to ROC-AUC 0.753 +/- 0.060 and Brier score 0.213. The
integrated logistic model achieved spatial ROC-AUC 0.756 +/- 0.092,
essentially matching XGBoost despite its simpler form (Figure 9; Table 3A). The approximately 0.17 AUC difference between random and spatial
validation therefore represents spatial optimism rather than
transferable model skill.

The built-environment block contributed more spatially transferable
discrimination than the vegetation block. Under spatial validation,
XGBoost achieved ROC-AUC 0.697 for M0, 0.722 for M1, 0.757 for M2, and
0.753 for M3. Logistic probabilities were reasonably calibrated overall,
with a mean recalibration slope of 1.02, whereas XGBoost was
overconfident, with a mean slope of 0.69 and visible deviation at
intermediate probabilities. Fold-to-fold variability was much larger
under spatial than random validation. Residual Moran' s I
remained 0.48-0.55, indicating that fire progression, ember transport,
suppression, and fine-scale exposure retained spatial structure not
captured by the open pre-fire predictors. The full metric set and
sensitivity specifications are reported in Table 3.

\begin{figure*}[H]
\centering
\includegraphics[width=\textwidth]{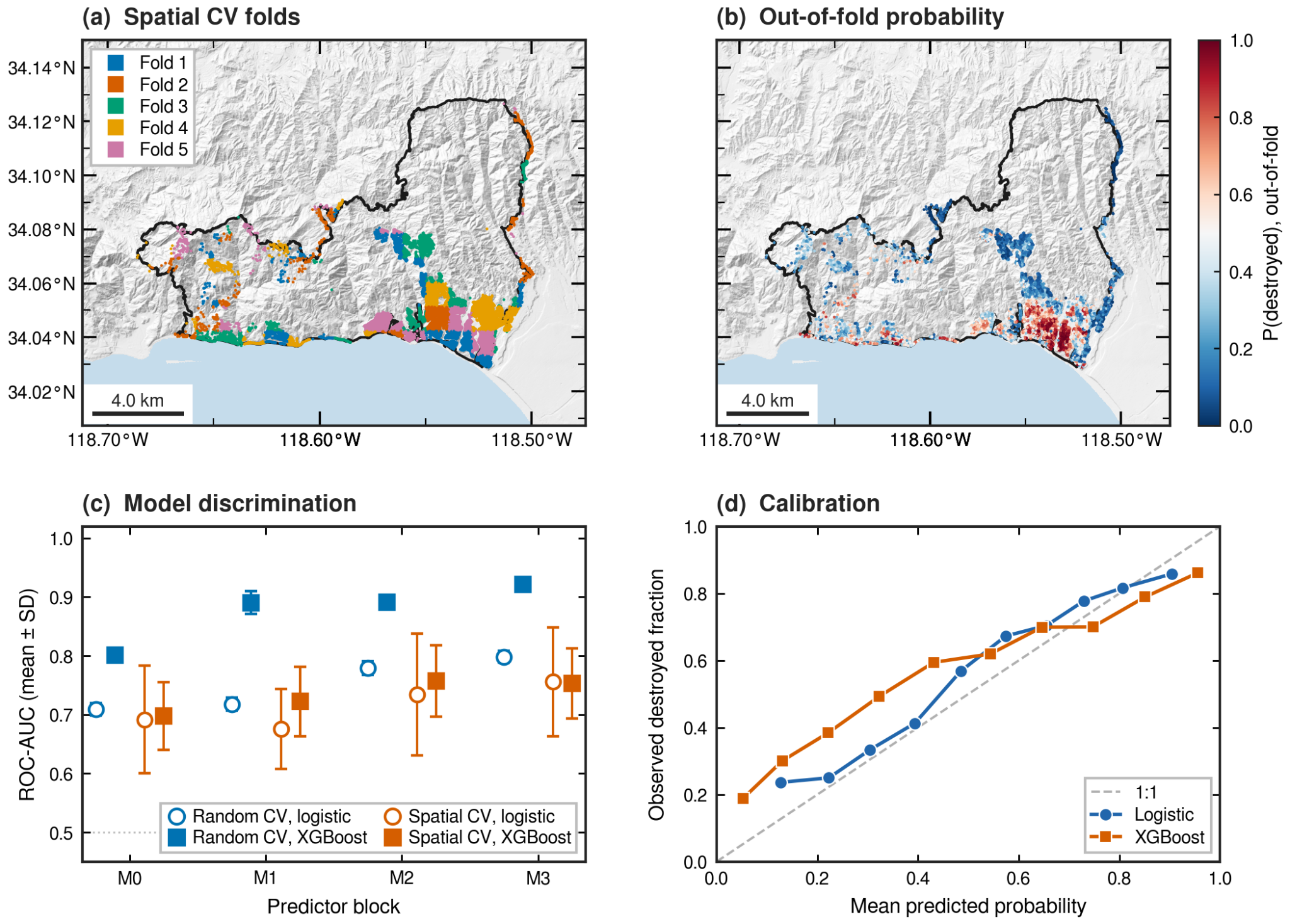}
\caption{Spatial validation and calibration. (A) One-kilometer spatial cross-validation folds; (B) out-of-fold destruction probability; (C) ROC-AUC by model block and validation scheme for logistic regression and XGBoost; and (D) calibration curves.}
\end{figure*}
\vspace{0.35\baselineskip}

\begin{table*}[t]
\centering
\textbf{Table 3.} Complete out-of-fold model performance and predeclared
sensitivity analyses. Values in parentheses are standard deviations
across folds.
\scriptsize
\setlength{\tabcolsep}{3pt}
\textbf{A. Model performance by validation scheme, predictor block, and
learner}
\vspace{0.15\baselineskip}
\begin{tabular*}{\textwidth}{@{\extracolsep{\fill}}>{\raggedright\arraybackslash}p{0.089\textwidth}>{\raggedright\arraybackslash}p{0.078\textwidth}>{\raggedright\arraybackslash}p{0.105\textwidth}>{\raggedright\arraybackslash}p{0.131\textwidth}>{\raggedright\arraybackslash}p{0.131\textwidth}>{\raggedright\arraybackslash}p{0.162\textwidth}>{\raggedright\arraybackslash}p{0.126\textwidth}>{\raggedright\arraybackslash}p{0.131\textwidth}@{}}
\toprule

\textbf{Validation}
 & \textbf{Block}
 & \textbf{Learner}
 & \textbf{ROC-AUC}
 & \textbf{PR-AUC}
 & \textbf{Balanced accuracy}
 & \textbf{F1}
 & \textbf{Brier}
 \\
\midrule
Random & M0 & Logistic & 0.709 (0.01) & 0.737 (0.014) & 0.662 (0.01) & 0.706 (0.011) & 0.217 (0.003) \\
 & M0 & XGBoost & 0.801 (0.004) & 0.834 (0.009) & 0.714 (0.006) & 0.754 (0.006) & 0.182 (0.001) \\
 & M1 & Logistic & 0.718 (0.011) & 0.744 (0.017) & 0.673 (0.01) & 0.72 (0.012) & 0.212 (0.004) \\
 & M1 & XGBoost & 0.891 (0.019) & 0.913 (0.018) & 0.804 (0.022) & 0.827 (0.018) & 0.135 (0.012) \\
 & M2 & Logistic & 0.779 (0.011) & 0.829 (0.009) & 0.719 (0.01) & 0.735 (0.009) & 0.192 (0.004) \\
 & M2 & XGBoost & 0.892 (0.009) & 0.917 (0.007) & 0.803 (0.007) & 0.818 (0.008) & 0.134 (0.006) \\
 & M3 & Logistic & 0.798 (0.01) & 0.843 (0.011) & 0.727 (0.008) & 0.747 (0.009) & 0.184 (0.004) \\
 & M3 & XGBoost & 0.921 (0.007) & 0.938 (0.005) & 0.842 (0.011) & 0.859 (0.008) & 0.112 (0.005) \\
Spatial & M0 & Logistic & 0.692 (0.091) & 0.702 (0.149) & 0.643 (0.059) & 0.667 (0.109) & 0.224 (0.029) \\
 & M0 & XGBoost & 0.697 (0.057) & 0.69 (0.115) & 0.625 (0.06) & 0.62 (0.12) & 0.234 (0.037) \\
 & M1 & Logistic & 0.676 (0.068) & 0.664 (0.122) & 0.641 (0.046) & 0.661 (0.095) & 0.227 (0.025) \\
 & M1 & XGBoost & 0.722 (0.059) & 0.718 (0.121) & 0.651 (0.049) & 0.644 (0.112) & 0.227 (0.032) \\
 & M2 & Logistic & 0.734 (0.103) & 0.745 (0.193) & 0.679 (0.095) & 0.652 (0.22) & 0.207 (0.034) \\
 & M2 & XGBoost & 0.757 (0.061) & 0.773 (0.118) & 0.679 (0.067) & 0.662 (0.14) & 0.211 (0.038) \\
 & M3 & Logistic & 0.756 (0.092) & 0.764 (0.165) & 0.698 (0.08) & 0.696 (0.163) & 0.199 (0.036) \\
 & M3 & XGBoost & 0.753 (0.06) & 0.769 (0.113) & 0.678 (0.066) & 0.661 (0.149) & 0.213 (0.042) \\
\bottomrule
\end{tabular*}
\vspace{0.35\baselineskip}
\textbf{B. Ring-scale, spatial-block, outcome, and population
sensitivity}
\vspace{0.15\baselineskip}
\begin{tabular*}{\textwidth}{@{\extracolsep{\fill}}>{\raggedright\arraybackslash}p{0.172\textwidth}>{\raggedright\arraybackslash}p{0.300\textwidth}>{\raggedright\arraybackslash}p{0.250\textwidth}>{\raggedright\arraybackslash}p{0.144\textwidth}>{\raggedright\arraybackslash}p{0.122\textwidth}@{}}
\toprule

\textbf{Dimension}
 & \textbf{Specification}
 & \textbf{Spatial-CV ROC-AUC (SD)}
 & \textbf{PR-AUC}
 & \textbf{N predictors}
 \\
\midrule
Ring scale & 0-30 m & 0.721 (0.073) & 0.74 & 21 \\
 & 30-100 m & 0.744 (0.060) & 0.768 & 22 \\
 & 100-300 m & 0.753 (0.042) & 0.767 & 22 \\
 & All scales & 0.756 (0.051) & 0.775 & 37 \\
Block size & 500 m & 0.809 (0.057) & 0.826 & 37 \\
 & 1000 m & 0.756 (0.051) & 0.775 & 37 \\
 & 2000 m & 0.765 (0.044) & 0.783 & 37 \\
Outcome/population & Destroyed+Major, residential & 0.765 (0.047) & 0.782 & 37 \\
 & Destroyed, all structures & 0.759 (0.053) & 0.735 & 37 \\
\bottomrule
\end{tabular*}
\end{table*}
\subsection{Dominant predictors and scale dependence}\label{dominant-predictors-and-scale-dependence}

Both model families identified neighborhood building arrangement as the
dominant signal (Figure 10; Table 4). In the logistic model, a
one-standard-deviation increase in building count within 100 m was
associated with a 4.12-fold increase in the odds of destruction (95\% CI 3.60-4.72). The 100 and 300 m building-count variables also dominated
XGBoost SHAP attributions. The SHAP response for building count within
100 m was nonlinear, with a pronounced increase above approximately
50-60 buildings. This value is a model-derived inflection within the
Palisades data, not a universal density threshold.

\begin{figure*}[H]
\centering
\includegraphics[width=\textwidth]{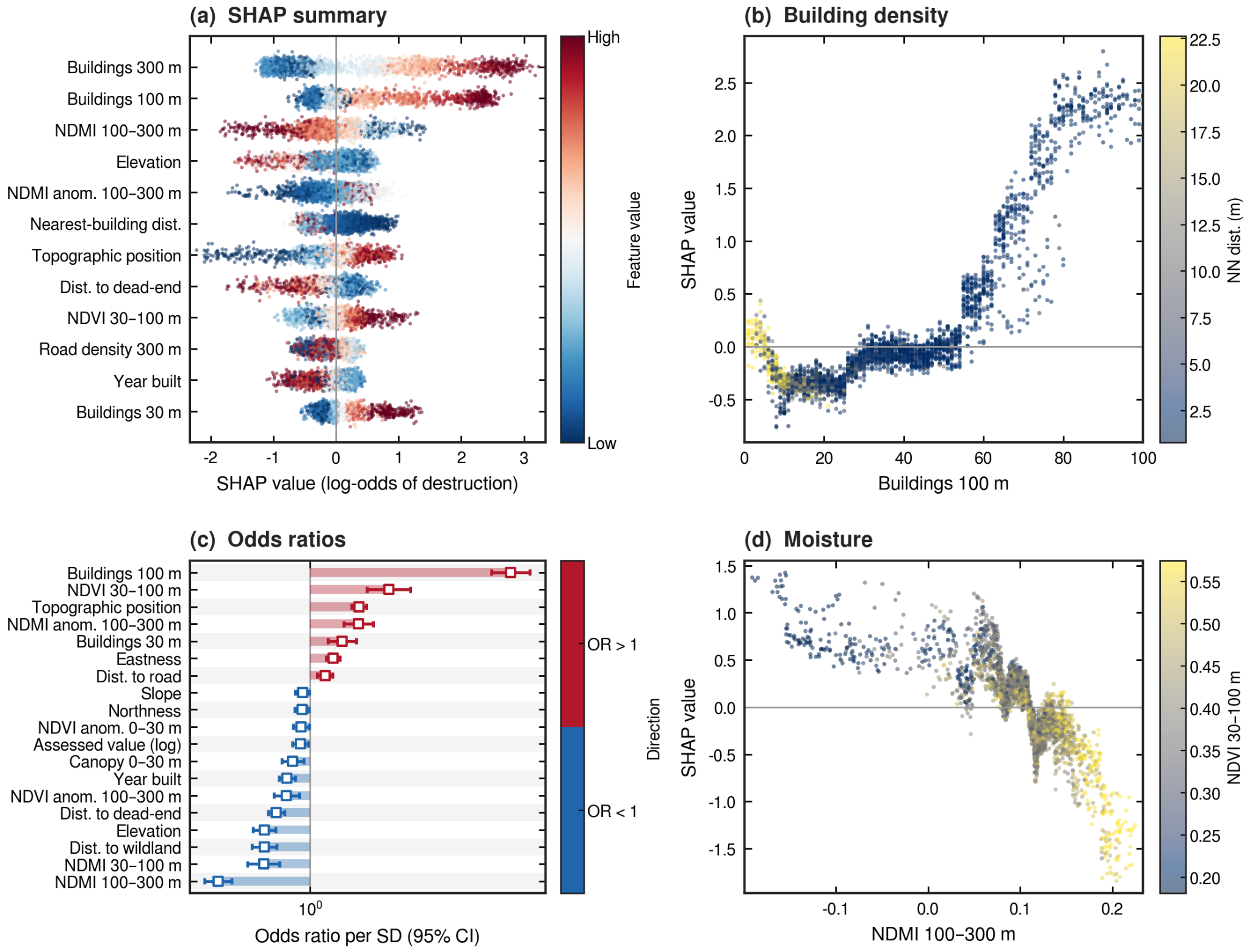}
\caption{Model interpretation. (A) SHAP summary for the twelve most influential XGBoost predictors; (B) SHAP dependence for building count within 100 m colored by nearest-neighbor distance; (C) standardized logistic-regression odds ratios with 95\% confidence intervals; and (D) SHAP dependence for NDMI in the 100-300 m ring colored by NDVI in the 30-100 m ring.}
\end{figure*}
\vspace{0.35\baselineskip}

\vspace{0.35\baselineskip}

\begin{table*}[t]
\centering
\textbf{Table 4.} Standardized logistic-regression associations and XGBoost
feature importance for the integrated predictor set (M3). Odds ratios
are reported per one-standard-deviation increase; mean
\textbar SHAP\textbar{} denotes mean absolute SHAP importance. Predictor
definitions and machine-readable codes are provided in Table 2.
\footnotesize
\setlength{\tabcolsep}{3pt}
\begin{tabular*}{\textwidth}{@{\extracolsep{\fill}}>{\raggedright\arraybackslash}p{0.316\textwidth}>{\raggedright\arraybackslash}p{0.141\textwidth}>{\raggedright\arraybackslash}p{0.202\textwidth}>{\raggedright\arraybackslash}p{0.094\textwidth}>{\raggedright\arraybackslash}p{0.121\textwidth}>{\raggedright\arraybackslash}p{0.101\textwidth}@{}}
\toprule

Predictor
 & OR per SD
 & 95\% CI
 & p
 & Mean \textbar SHAP\textbar{}
 & SHAP rank
 \\
\midrule
Buildings within 100 m & 4.123 & (3.601, 4.722) & \textless0.001 & 0.464
& 2 \\
Pre-fire NDVI, 30-100 m & 1.743 & (1.495, 2.031) & \textless0.001 &
0.235 & 9 \\
Topographic position index, 300 m & 1.414 & (1.342, 1.490) &
\textless0.001 & 0.278 & 7 \\
NDMI anomaly, 100-300 m & 1.406 & (1.267, 1.561) & \textless0.001 &
0.306 & 5 \\
Buildings within 30 m & 1.253 & (1.135, 1.383) & \textless0.001 & 0.203
& 12 \\
Eastness & 1.175 & (1.120, 1.233) & \textless0.001 & 0.135 & 21 \\
Distance to nearest road & 1.112 & (1.054, 1.173) & \textless0.001 &
0.080 & 31 \\
Pre-fire NDVI, 0-30 m & 1.098 & (0.982, 1.229) & 0.101 & 0.199 & 13 \\
Pre-fire NDMI, 0-30 m & 1.066 & (0.977, 1.163) & 0.149 & 0.069 & 34 \\
Pre-fire LST, 30-100 m & 1.056 & (0.995, 1.120) & 0.074 & 0.196 & 14 \\
Nearest-building distance & 1.051 & (0.994, 1.111) & 0.078 & 0.289 &
6 \\
NDMI anomaly, 30-100 m & 1.045 & (0.968, 1.127) & 0.259 & 0.100 & 26 \\
Canopy cover, 30-100 m & 1.034 & (0.954, 1.121) & 0.419 & 0.060 & 35 \\
Road density within 300 m & 1.034 & (0.914, 1.169) & 0.600 & 0.211 &
10 \\
Wildland fuel fraction, 0-30 m & 1.032 & (0.952, 1.118) & 0.449 & 0.017
& 37 \\
Wildland fuel fraction, 100-300 m & 0.992 & (0.880, 1.118) & 0.894 &
0.135 & 20 \\
NDMI anomaly, 0-30 m & 0.986 & (0.927, 1.048) & 0.643 & 0.073 & 33 \\
NDVI anomaly, 30-100 m & 0.980 & (0.911, 1.055) & 0.595 & 0.100 & 27 \\
Building footprint area & 0.975 & (0.919, 1.035) & 0.406 & 0.150 & 18 \\
Slope & 0.947 & (0.901, 0.995) & 0.031 & 0.088 & 30 \\
Northness & 0.943 & (0.897, 0.992) & 0.022 & 0.076 & 32 \\
NDVI anomaly, 0-30 m & 0.939 & (0.885, 0.996) & 0.035 & 0.127 & 25 \\
Assessed improvement value (log10) & 0.932 & (0.882, 0.985) & 0.013 &
0.128 & 24 \\
Canopy cover, 0-30 m & 0.884 & (0.818, 0.954) & 0.002 & 0.027 & 36 \\
Year built & 0.850 & (0.802, 0.901) & \textless0.001 & 0.205 & 11 \\
NDVI anomaly, 100-300 m & 0.846 & (0.774, 0.925) & \textless0.001 &
0.094 & 29 \\
Distance to nearest dead-end node & 0.787 & (0.744, 0.834) &
\textless0.001 & 0.250 & 8 \\
Elevation & 0.723 & (0.667, 0.783) & \textless0.001 & 0.339 & 4 \\
Distance to nearest wildland fuel & 0.722 & (0.662, 0.788) &
\textless0.001 & 0.144 & 19 \\
Pre-fire NDMI, 30-100 m & 0.720 & (0.642, 0.806) & \textless0.001 &
0.158 & 17 \\
Pre-fire NDMI, 100-300 m & 0.522 & (0.474, 0.575) & \textless0.001 &
0.352 & 3 \\
\bottomrule
\end{tabular*}
\vspace{0.15\baselineskip}
\parbox[t]{\textwidth}{\footnotesize Note. Odds ratios (ORs) are reported per one-standard-deviation increase
in each continuous predictor, with 95\% confidence intervals and
model-based p-values. Mean \textbar SHAP\textbar{} denotes mean absolute
SHAP value and quantifies global XGBoost feature importance; SHAP rank
refers to the integrated predictor set. Predictor definitions,
machine-readable variable names, data sources, and spatial support are
provided in Table 2. Intervals and p-values should be interpreted in
light of the residual spatial dependence reported in the text.}
\end{table*}
Vegetation amount and moisture had opposing conditional relationships.
NDMI at 100-300 m was the strongest protective environmental variable
(odds ratio 0.52, 95\% CI 0.47-0.58), and NDMI at 30-100 m acted in the
same direction (0.72). After moisture was held constant, NDVI at 30-100
m was positively associated with destruction (1.74, 1.50-2.03).
Immediate-ring canopy cover was weakly protective (0.88, 0.82-0.95). The
positive coefficient for the 100-300 m NDMI anomaly occurred alongside a
strongly protective NDMI level and is interpreted as partial suppression
among correlated level and anomaly terms, not as an independent
hazardous moisture anomaly.

Terrain, access, and construction-era effects were smaller but coherent.
Destruction odds increased at higher local topographic positions and on
east-facing aspects and decreased with elevation, distance from wildland
fuels, and distance from dead-end nodes. Newer construction and greater
assessed value were protective conditional on the other predictors.
Restricting vegetation and density variables to a single ring showed
that spatially validated information increased with scale: ROC-AUC was
0.721 at 0-30 m, 0.744 at 30-100 m, and 0.753 at 100-300 m, compared
with 0.756 for the full multi-scale model (Figure 11). The 100-300 m
ring also carried the largest aggregate SHAP contribution. Results were
stable when Major damage was grouped with Destroyed, when all structure
categories were included, and when 2 km blocks were used.
Five-hundred-meter blocks produced higher AUC (0.809), consistent with
greater proximity between training and test observations. Scale,
block-size, outcome, and population sensitivity results are reported in
Figure 11 and Table 3B.

\begin{figure*}[H]
\centering
\includegraphics[width=\textwidth]{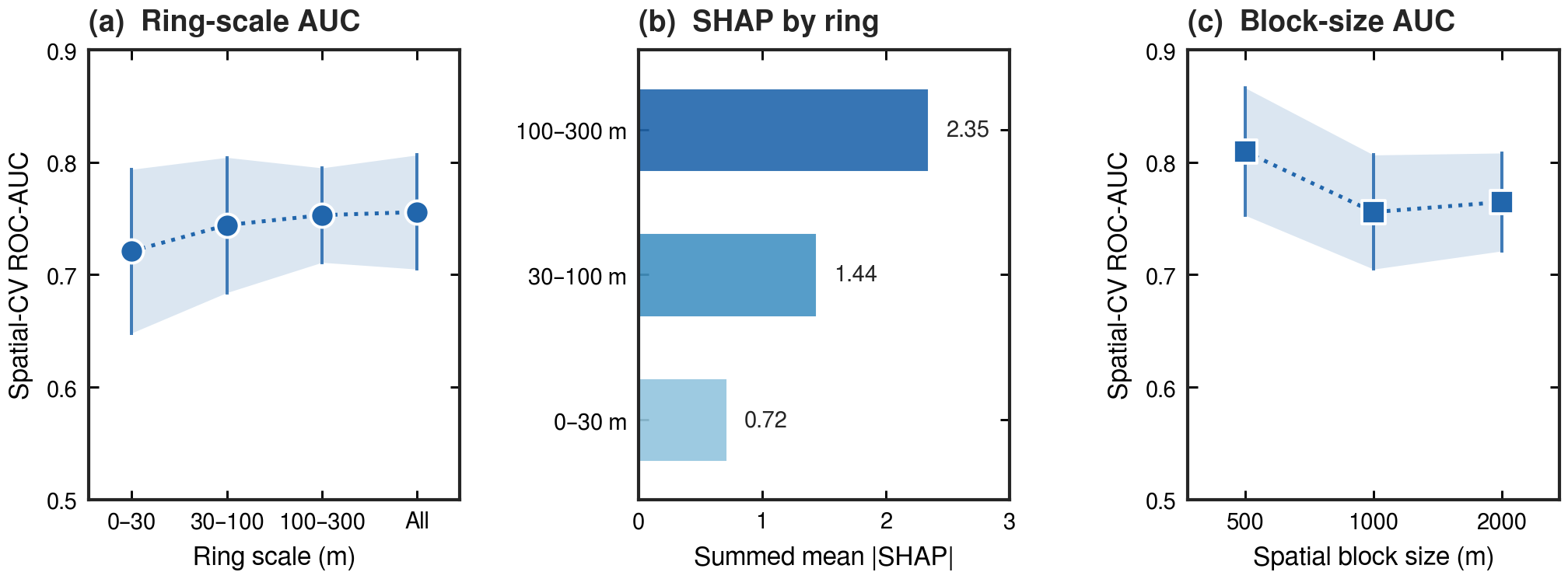}
\caption{Scale and block-size sensitivity. (A) Spatial cross-validation ROC-AUC by ring scale; (B) summed mean absolute SHAP contribution by ring; and (C) spatial-block-size sensitivity of ROC-AUC.}
\end{figure*}

\subsection{Fire impact and early ecological response}\label{fire-impact-and-early-ecological-response}

\begin{figure*}[H]
\centering
\includegraphics[width=\textwidth]{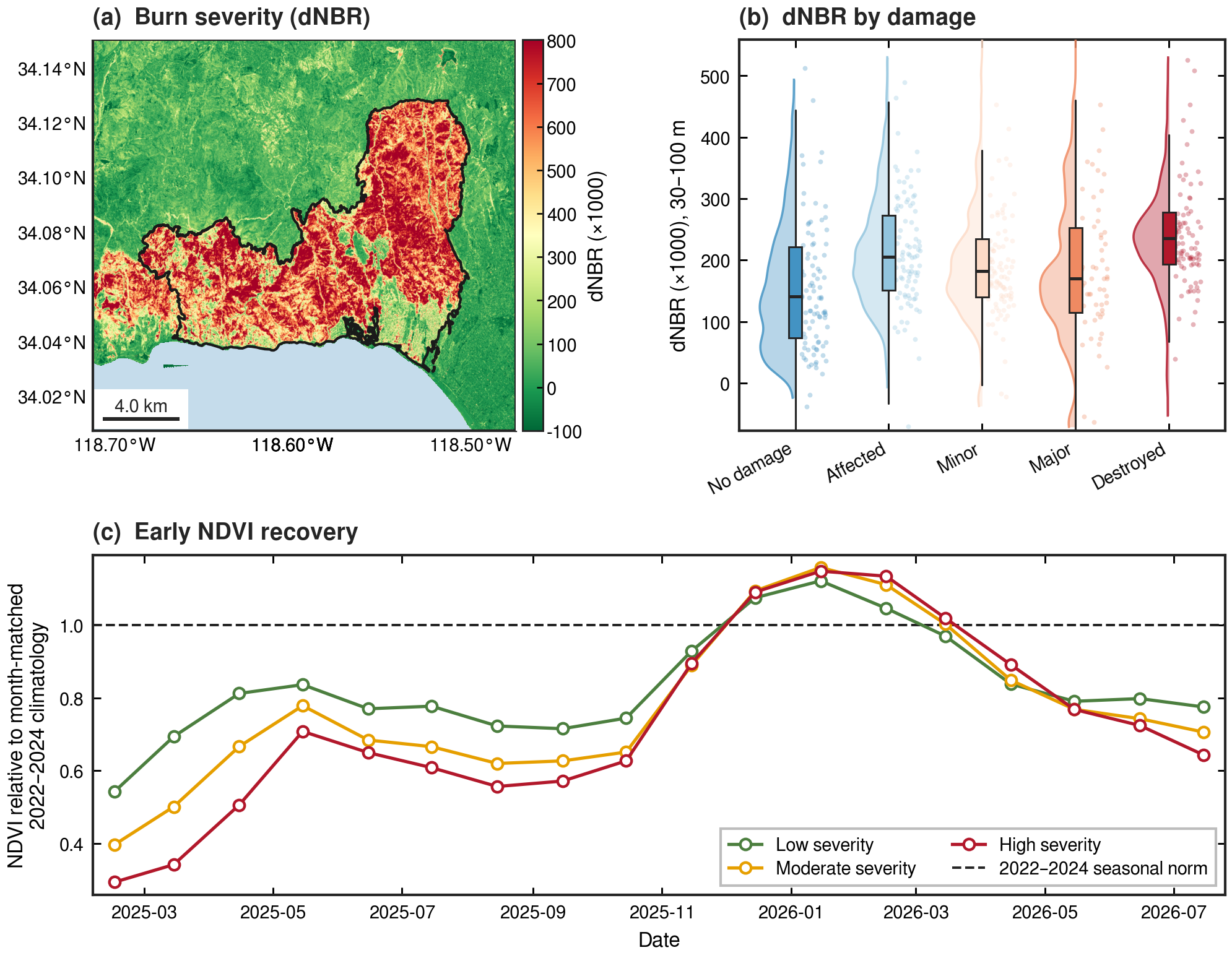}
\caption{Fire impact and early vegetation response. (A) Study-derived dNBR; (B) dNBR within the 30-100 m ring by DINS damage class; and (C) monthly NDVI recovery by burn-severity class relative to month-matched 2022-2024 climatology from February 2025 through July 2026.}
\end{figure*}

Study-derived dNBR classified 71\% of the burned perimeter (68 km2) as
high severity (Figure 12). Mean dNBR in the 30-100 m ring increased from
159 for structures with no recorded damage to 218 for Affected and 236
for Destroyed structures, although the broad overlap among classes
confirms that 10 m spectral change does not resolve structure-scale
exposure (Figure 12).

\vspace{0.35\baselineskip}

Vegetation response was severity ordered and strongly seasonal. In the
first post-fire month, NDVI was 54\%, 40\%, and 29\% of the
month-matched norm in low-, moderate-, and high-severity areas,
respectively. All classes approached or briefly exceeded the seasonal
norm during the wet winter of 2025-2026, consistent with an herbaceous
post-fire flush, but fell to 64-77\% of normal by July 2026. Early
greenness recovery therefore did not represent recovery of pre-fire
vegetation structure. The full severity and recovery sequence is shown
in Figure 12.

\subsection{Community context}\label{community-context}

\begin{figure*}[H]
\centering
\includegraphics[width=\textwidth]{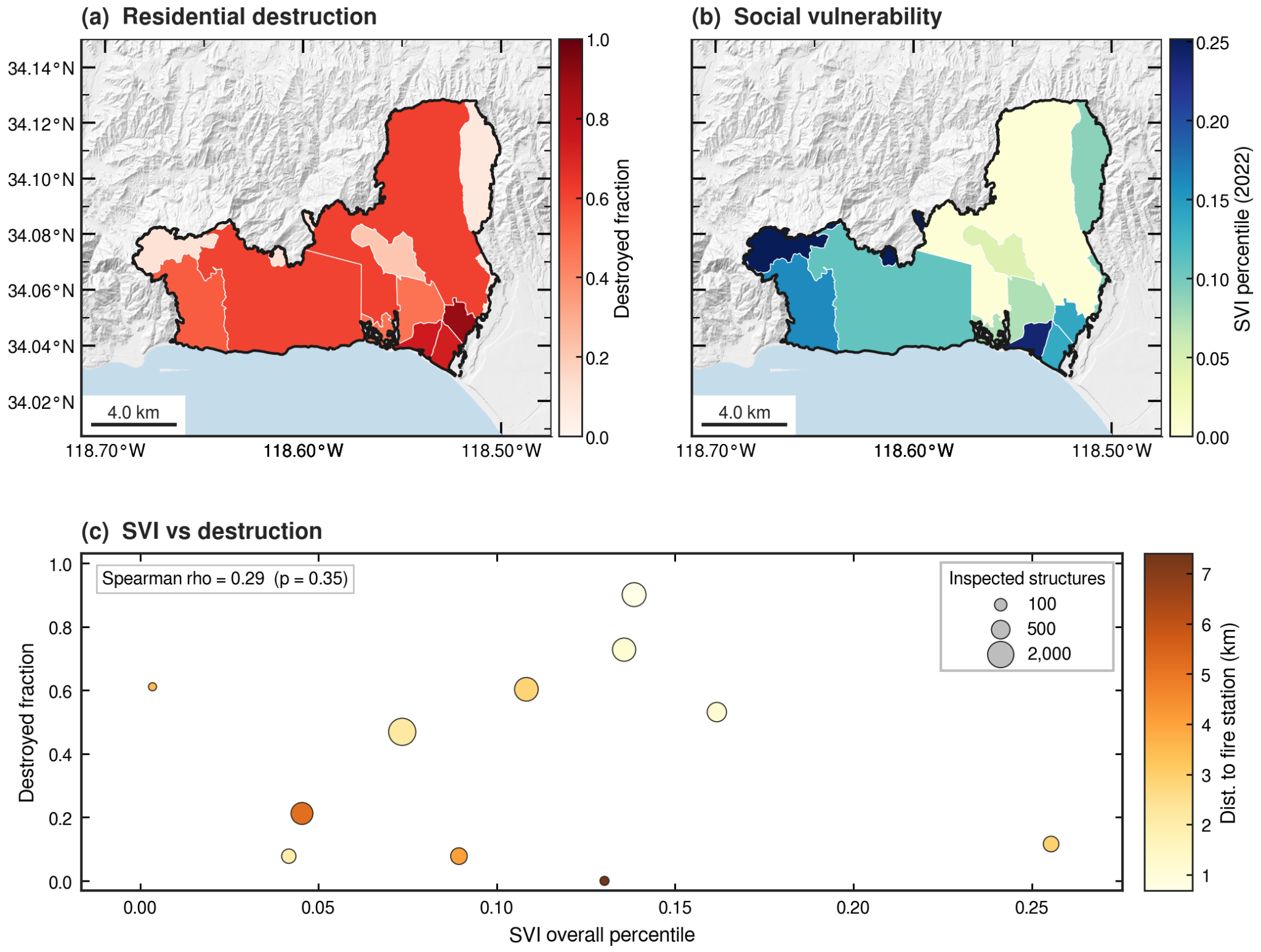}
\caption{Community context. (A) Residential destruction fraction by census tract; (B) CDC/ATSDR SVI 2022 overall percentile by tract; and (C) tract-level SVI versus residential destruction. Point size represents the number of inspected structures and point color represents median network distance to the nearest fire station.}
\end{figure*}

The twelve census tracts containing inspected structures occupied a
narrow and low social-vulnerability range: SVI overall percentiles were
0.00-0.26, all within the least-vulnerable national quartile. Tract
destruction fractions ranged from 0 to 0.90, but their association with
SVI was not detectable (Spearman rho = 0.29, p = 0.35; Figure 13).
Median network distance to the nearest fire station ranged from 0.7 to
7.4 km without a consistent destruction gradient. This null result does
not imply that social vulnerability is unimportant to wildfire
resilience; it shows that this single, affluent study area contains too
little social variation to estimate that relationship reliably \citep{davies2018,norlen2026}.

\vspace{0.35\baselineskip}

\section{Discussion}\label{discussion}

\subsection{Neighborhood morphology as a resilience mechanism}\label{neighborhood-morphology-as-a-resilience-mechanism}

The clearest result is that urban morphology, especially neighborhood
building concentration, carried more transferable information about
destruction than any single environmental variable. This independently
corroborates the building-density findings of \citet{kenny2026} and
the multi-scale urban-morphology results of \citet{norlen2026}, while
extending them in three ways: the association is estimated conditional
on a full pre-fire environmental stack, it is evaluated outside local
spatial blocks, and its probability calibration is reported. The
nonlinear response above roughly 50-60 buildings within 100 m is
consistent with a change in propagation opportunity as repeated ember
and radiant-heat exposure accumulates across dense fabric. Because the
analysis is observational, the inflection should be used to identify
neighborhoods requiring closer assessment, not as a zoning cutoff or
causal density rule.

The result also clarifies the scale at which open urban data are most
useful. Nearest-neighbor distance was short almost everywhere, whereas
building counts across 100-300 m distinguished urban fabric with very
different loss rates. That neighborhood support captures the cumulative
exposure created by many possible ignition pathways, a pattern
consistent with prior California structure-loss studies and the Camp
Fire analysis \citep{syphard2012,knapp2021,syphard2021}. The protective year-built association is compatible with
progressive improvements in construction practice, but year built also
encodes neighborhood age, layout, and redevelopment history. It should
therefore guide screening for retrofit needs rather than be interpreted
as a direct building-code effect.

\subsection{Vegetation condition reframes the greenness-risk relationship}\label{vegetation-condition-reframes-the-greenness-risk-relationship}

The environmental findings argue against treating urban vegetation as
uniformly protective or hazardous. At a given vegetation amount, greater
neighborhood moisture was strongly protective; at a given moisture
level, greater vegetation amount was associated with higher destruction
odds. In the exceptionally dry months preceding the fire, this
combination is physically plausible: abundant vegetation can represent
combustible biomass, whereas moisture limits ignitability and fire
spread. The immediate-ring canopy effect was small and partly
protective, further weakening any interpretation that simple canopy
removal would improve resilience.

For urban greening and defensible-space policy, the actionable
distinction is therefore among vegetation amount, moisture condition,
arrangement, and continuity. NDVI alone cannot make that distinction.
Moisture-sensitive indices, irrigation context, fuel type, and
maintenance need to be interpreted together, especially at the 100-300 m
neighborhood scale where the transferable signal was strongest \citep{dennison2005,escobedo2025}. Species-specific attention models
likewise show that NIR and SWIR responses depend on interacting
structural, biochemical, and water-related traits, cautioning against
universal spectral interpretations \citep{farajpoor2025b}.
Higher-resolution canopy and parcel data could refine local
implementation; recent open optical GeoAI work demonstrates how
updateable canopy screening can complement structural inventories while
retaining spatial uncertainty \citep{narimani2026b}.

\subsection{Spatial validation is a substantive urban-informatics result}\label{spatial-validation-is-a-substantive-urban-informatics-result}

The 0.17 AUC optimism gap is not a technical footnote. It is the
difference between an apparently excellent model and a moderately
informative neighborhood-screening model. Gradient boosting benefited
most from random spatial mixing, yet it did not outperform logistic
regression once neighborhoods were held out. Model complexity therefore
added little transferable value in this single-event setting.
Calibration moved in the same direction: XGBoost was overconfident
outside local blocks, while the logistic model remained closer to
observed frequencies.

This result has direct implications for urban informatics. A
data-to-decision system is credible only if evaluation resembles the
decision it will support. Randomly withholding nearby structures
evaluates interpolation within familiar neighborhoods; spatial blocking
more closely approximates prioritization in a different part of the
city. Residual autocorrelation shows that even block validation cannot
reproduce omitted fire dynamics. Accordingly, the out-of-fold surface
should be interpreted as a calibrated susceptibility screen, not a
deterministic parcel forecast. Spatially structured validation,
calibration, and block-size sensitivity should be minimum reporting
elements for single-event urban hazard models \citep{roberts2017,ploton2020}. Related geospatial-embedding work similarly found
that predictive uncertainty concentrates near class boundaries,
reinforcing the value of spatially independent evaluation and explicit
uncertainty surfaces \citep{narimani2026c}.

\subsection{Translating diagnostics into planning and design}\label{translating-diagnostics-into-planning-and-design}

The workflow supports three practical uses. First, rebuilding and
retrofit programs can use neighborhood building concentration to
identify places where hardening only an isolated structure is unlikely
to be sufficient. The relevant intervention unit is the
adjacent-building interface and surrounding block: ember-resistant vents
and roofs, ignition-resistant fencing, reduced combustible continuity,
and coordinated treatment of the closest structures. Second, vegetation
programs can replace greenness-only screening with moisture-aware
assessments that preserve the cooling and ecological value of urban
canopy while targeting dry, continuous fuels. Third, agencies can use
the spatially validated probability surface and its calibration to
prioritize field inspection, not to automate enforcement or insurance
decisions.

The key design principle is to couple every indicator with its valid
scale and uncertainty. This aligns with decision-oriented urban
informatics, in which open data, computing, sensing, and governance are
connected through an explicit implementation pathway rather than ending
with a map \citep{shi2022,pradeep2026}. High-resolution farmland
segmentation has likewise shown why scene-separated evaluation and
explicit boundary uncertainty are essential when mapped outputs are
intended for operational use \citep{narimani2026a}. Because the
acquisition and analysis are scripted, the same workflow can be rerun
shortly after damage-inspection data are released, creating comparable
evidence across events and supporting a future multi-city benchmark.
Such benchmarking is essential before any Palisades-specific threshold
is generalized.

\subsection{Scope, transferability, and unresolved processes}\label{scope-transferability-and-unresolved-processes}

The design intentionally holds one extreme event fixed and asks why
inspected structures within that event experienced different outcomes.
This strengthens the analysis of spatial contrasts but does not estimate
how changing wind, ignition location, or suppression would alter loss
across events. Inference is conditional on DINS-inspected structures,
and structures outside that frame remain unknown. Building-envelope
characteristics were not modeled because their post-fire observability
was strongly outcome dependent; excluding them protects internal
validity while leaving a known component of home ignitability unresolved
\citep{zamanialaei2025}.

Sensor support defines a second boundary. Ten- to 30 m products
characterize local and neighborhood context but cannot resolve
individual yard treatments, fence materials, vents, or fine fuel
continuity. The value of matching sensor support to the process of
interest is also evident in UAV stress-detection studies, where
centimeter-scale observations resolve canopy heterogeneity that is
necessarily averaged in satellite products \citep{narimani2024}. The
strong 100-300 m signal should therefore be understood as the scale at
which open data transfer best in this study, not as evidence that
parcel-level conditions are irrelevant. Similarly, dNBR combines
vegetation and structural change in urban areas and is retained only as
an impact descriptor. The SVI result is restricted by twelve
low-vulnerability tracts and should motivate cross-event comparison with
more diverse communities rather than a general conclusion about equity.

These boundaries point directly to the next analytical step: replicate
the pipeline across DINS events, add independent structure-hardening
data, incorporate time-resolved fire progression and wind exposure, and
evaluate whether neighborhood effects remain stable across urban forms
and social contexts. The present study supplies the open baseline needed
for that comparison. Its contribution is not an all-purpose wildfire
forecast, but a transparent account of what public urban data can
resolve, where they remain uncertain, and how those limits should shape
planning use.

\section{Conclusion}\label{conclusion}

A reproducible urban-informatics analysis of 12,081 Palisades Fire
damage inspections shows that neighborhood structure and vegetation
moisture were the most informative pre-fire correlates of residential
destruction. Building count within 100 m produced the largest
association and a marked nonlinear model response, while
moisture-sensitive vegetation information was protective and greenness
alone was not. These relationships were visible only within an
evaluation framework that respected space: the integrated
gradient-boosting model fell from ROC-AUC 0.92 under random
cross-validation to approximately 0.75 under 1 km spatial blocking,
where a simpler logistic model performed equally well and was better
calibrated. The practical outcome is a neighborhood-scale screening
framework, not a parcel-level prediction system. It can support
coordinated hardening of dense urban fabric, moisture-aware vegetation
management, and targeted field assessment while communicating
uncertainty explicitly. The separated severity, recovery, and community
analyses broaden the resilience interpretation without contaminating the
pre-fire model. By combining open data, temporal leakage control,
spatial validation, transparent model interpretation, and a direct
planning translation, the workflow provides a transferable basis for
comparative urban wildfire resilience studies.

\section*{Data availability statement}\label{data-availability-statement}

All public source products and collection identifiers are listed in
Table 1, and persistent source links, retrieval metadata, and checksums
are documented in the repository manifest. The analysis-ready
structure-level feature table, study boundary, manuscript tables, model
diagnostics, out-of-fold predictions, SHAP sample, final figures,
manifests, and related derived products are openly available in Zenodo
at \url{https://doi.org/10.5281/zenodo.22061862} \citep{palisades2026data}. The same record provides the clipped Sentinel-2, Landsat,
LANDFIRE, terrain, and supporting census inputs used to rebuild the
predictors, subject to the licenses of the original providers.

\section*{Code availability statement}\label{code-availability-statement}

Replication code, configuration files, environment specifications,
automated tests, and reproducibility documentation are openly available
at
\url{https://github.com/MohammadrezaNarimaniUCDavis/Palisades_Urban_Wildfire_GeoAI}.
The repository includes the scripted workflow used to rebuild the
principal tables, model diagnostics, and figures from the archived data
products.

\section*{Ethics statement}\label{ethics-statement}

The study used publicly available geospatial, environmental,
infrastructure, census, and structure-inspection data. It did not
involve human participants, identifiable personal information, animals,
or an intervention; institutional ethics approval and informed consent
were therefore not required.

\section*{Author contributions}\label{author-contributions}

Parastoo Farajpoor: Conceptualization, Methodology, Validation,
Visualization, Writing - original draft, Writing - review and editing.
Mohammadreza Narimani: Conceptualization, Data curation, Formal
analysis, Methodology, Software, Validation, Visualization, Writing -
original draft, Writing - review and editing, Project administration.
Both authors approved the submitted version and accept accountability
for the work.

\section*{Funding}\label{funding}

The authors declare that no external financial support was reported for
the research, authorship, and/or publication of this article.

\section*{Conflict of interest}\label{conflict-of-interest}

The authors declare that the research was conducted in the absence of
any commercial or financial relationships that could be construed as a
potential conflict of interest.

\section*{Generative AI statement}\label{generative-ai-statement}

During the preparation of this work, the authors used ChatGPT to improve
grammatical accuracy, refine sentence structure, and enhance
visualizations. All AI-generated revisions were thoroughly reviewed and
edited by the authors to ensure relevance and accuracy.

\section*{Acknowledgments}\label{acknowledgments}

The authors acknowledge CAL FIRE, the National Interagency Fire Center,
the U.S. Geological Survey, the U.S. Department of Agriculture,
Copernicus, ECMWF, the Climatology Lab, the U.S. Census Bureau,
CDC/ATSDR, and OpenStreetMap contributors for maintaining the public
data resources used in this study. This study contains modified
Copernicus Sentinel data. OpenStreetMap data are available under the
Open Database License.

\section*{Abbreviations}\label{abbreviations}

AUC, area under the curve; DINS, Damage Inspection; dNBR, differenced
Normalized Burn Ratio; GEE, Google Earth Engine; NDMI, Normalized
Difference Moisture Index; NDVI, Normalized Difference Vegetation Index;
OSM, OpenStreetMap; SHAP, Shapley additive explanations; SVI, Social
Vulnerability Index; WUI, wildland-urban interface.

\bibliographystyle{IEEEtranN}
\bibliography{references,arxiv_supplement}

% Generated by IEEEtranN.bst, version: 1.14 (2015/08/26)
\begin{thebibliography}{59}
\providecommand{\natexlab}[1]{#1}
\providecommand{\url}[1]{#1}
\csname url@samestyle\endcsname
\providecommand{\newblock}{\relax}
\providecommand{\bibinfo}[2]{#2}
\providecommand{\BIBentrySTDinterwordspacing}{\spaceskip=0pt\relax}
\providecommand{\BIBentryALTinterwordstretchfactor}{4}
\providecommand{\BIBentryALTinterwordspacing}{\spaceskip=\fontdimen2\font plus
\BIBentryALTinterwordstretchfactor\fontdimen3\font minus
  \fontdimen4\font\relax}
\providecommand{\BIBforeignlanguage}[2]{{%
\expandafter\ifx\csname l@#1\endcsname\relax
\typeout{** WARNING: IEEEtranN.bst: No hyphenation pattern has been}%
\typeout{** loaded for the language `#1'. Using the pattern for}%
\typeout{** the default language instead.}%
\else
\language=\csname l@#1\endcsname
\fi
#2}}
\providecommand{\BIBdecl}{\relax}
\BIBdecl

\bibitem[Cohen(2000)]{cohen2000}
J.~D. Cohen, ``Preventing disaster: Home ignitability in the wildland-urban
  interface,'' \emph{Journal of Forestry}, vol.~98, no.~3, pp. 15--21, 2000.

\bibitem[Calkin et~al.(2023)Calkin, Barrett, Cohen, Finney, Pyne, and
  Quarles]{calkin2023}
D.~E. Calkin, K.~Barrett, J.~D. Cohen, M.~A. Finney, S.~J. Pyne, and S.~L.
  Quarles, ``Wildland-urban fire disasters aren’t actually a wildfire
  problem,'' \emph{Proceedings of the National Academy of Sciences}, vol. 120,
  no.~51, p. e2315797120, 2023.

\bibitem[Metz et~al.(2024)Metz, Fischer, and Liel]{metz2024}
A.~J. Metz, E.~C. Fischer, and A.~B. Liel, ``The influence of housing, parcel,
  and neighborhood characteristics on housing survival in the marshall fire,''
  \emph{Fire Technology}, vol.~60, no.~6, pp. 4065--4097, 2024.

\bibitem[Meerow et~al.(2016)Meerow, Newell, and Stults]{meerow2016}
S.~Meerow, J.~P. Newell, and M.~Stults, ``Defining urban resilience: A
  review,'' \emph{Landscape and Urban Planning}, vol. 147, pp. 38--49, 2016.

\bibitem[McWethy et~al.(2019)McWethy, Schoennagel, Higuera, Krawchuk, Harvey,
  Metcalf, Schultz, Miller, Metcalf, Buma, Virapongse, Kulig, Stedman,
  Ratajczak, Nelson, and Kolden]{mcwethy2019}
D.~B. McWethy, T.~Schoennagel, P.~E. Higuera, M.~Krawchuk, B.~J. Harvey, E.~C.
  Metcalf, C.~Schultz, C.~Miller, A.~L. Metcalf, B.~Buma, A.~Virapongse, J.~C.
  Kulig, R.~C. Stedman, Z.~Ratajczak, C.~R. Nelson, and C.~Kolden, ``Rethinking
  resilience to wildfire,'' \emph{Nature Sustainability}, vol.~2, no.~9, pp.
  797--804, 2019.

\bibitem[Radeloff et~al.(2018)Radeloff, Helmers, Kramer, Mockrin, Alexandre,
  Bar-Massada, Butsic, Hawbaker, Martinuzzi, Syphard, and
  Stewart]{radeloff2018}
V.~C. Radeloff, D.~P. Helmers, H.~A. Kramer, M.~H. Mockrin, P.~M. Alexandre,
  A.~Bar-Massada, V.~Butsic, T.~J. Hawbaker, S.~Martinuzzi, A.~D. Syphard, and
  S.~I. Stewart, ``Rapid growth of the us wildland-urban interface raises
  wildfire risk,'' \emph{Proceedings of the National Academy of Sciences}, vol.
  115, no.~13, pp. 3314--3319, 2018.

\bibitem[Carlson et~al.(2022)Carlson, Helmers, Hawbaker, Mockrin, and
  Radeloff]{carlson2022}
A.~R. Carlson, D.~P. Helmers, T.~J. Hawbaker, M.~H. Mockrin, and V.~C.
  Radeloff, ``The wildland–urban interface in the united states based on 125
  million building locations,'' \emph{Ecological Applications}, vol.~32, no.~5,
  p. e2597, 2022.

\bibitem[Schug et~al.(2023)Schug, Bar-Massada, Carlson, Cox, Hawbaker, Helmers,
  Hostert, Kaim, Kasraee, Martinuzzi, Mockrin, Pfoch, and Radeloff]{schug2023}
F.~Schug, A.~Bar-Massada, A.~R. Carlson, H.~Cox, T.~J. Hawbaker, D.~Helmers,
  P.~Hostert, D.~Kaim, N.~K. Kasraee, S.~Martinuzzi, M.~H. Mockrin, K.~A.
  Pfoch, and V.~C. Radeloff, ``The global wildland–urban interface,''
  \emph{Nature}, vol. 621, no. 7977, pp. 94--99, 2023.

\bibitem[Syphard et~al.(2012)Syphard, Keeley, Massada, Brennan, and
  Radeloff]{syphard2012}
A.~D. Syphard, J.~E. Keeley, A.~B. Massada, T.~J. Brennan, and V.~C. Radeloff,
  ``Housing arrangement and location determine the likelihood of housing loss
  due to wildfire,'' \emph{PLoS ONE}, vol.~7, no.~3, p. e33954, 2012.

\bibitem[Kramer et~al.(2019)Kramer, Mockrin, Alexandre, and
  Radeloff]{kramer2019}
H.~A. Kramer, M.~H. Mockrin, P.~M. Alexandre, and V.~C. Radeloff, ``High
  wildfire damage in interface communities in california,'' \emph{International
  Journal of Wildland Fire}, vol.~28, no.~9, pp. 641--650, 2019.

\bibitem[Syphard and Keeley(2019)]{syphard2019}
A.~Syphard and J.~Keeley, ``Factors associated with structure loss in the
  2013–2018 california wildfires,'' \emph{Fire}, vol.~2, no.~3, p.~49, 2019.

\bibitem[Calkin et~al.(2014)Calkin, Cohen, Finney, and Thompson]{calkin2014}
D.~E. Calkin, J.~D. Cohen, M.~A. Finney, and M.~P. Thompson, ``How risk
  management can prevent future wildfire disasters in the wildland-urban
  interface,'' \emph{Proceedings of the National Academy of Sciences}, vol.
  111, no.~2, pp. 746--751, 2014.

\bibitem[Syphard et~al.(2014)Syphard, Brennan, and Keeley]{syphard2014}
A.~D. Syphard, T.~J. Brennan, and J.~E. Keeley, ``The role of defensible space
  for residential structure protection during wildfires,'' \emph{International
  Journal of Wildland Fire}, vol.~23, no.~8, pp. 1165--1175, 2014.

\bibitem[Knapp et~al.(2021)Knapp, Valachovic, Quarles, and Johnson]{knapp2021}
E.~E. Knapp, Y.~S. Valachovic, S.~L. Quarles, and N.~G. Johnson, ``Housing
  arrangement and vegetation factors associated with single-family home
  survival in the 2018 camp fire, california,'' \emph{Fire Ecology}, vol.~17,
  no.~1, p.~25, 2021.

\bibitem[Syphard et~al.(2021)Syphard, Rustigian-Romsos, and
  Keeley]{syphard2021}
A.~D. Syphard, H.~Rustigian-Romsos, and J.~E. Keeley, ``Multiple-scale
  relationships between vegetation, the wildland–urban interface, and
  structure loss to wildfire in california,'' \emph{Fire}, vol.~4, no.~1,
  p.~12, 2021.

\bibitem[Dennison et~al.(2005)Dennison, Roberts, Peterson, and
  Rechel]{dennison2005}
P.~E. Dennison, D.~A. Roberts, S.~H. Peterson, and J.~Rechel, ``Use of
  normalized difference water index for monitoring live fuel moisture,''
  \emph{International Journal of Remote Sensing}, vol.~26, no.~5, pp.
  1035--1042, 2005.

\bibitem[Escobedo et~al.(2025)Escobedo, Yadav, Cappelluti, and
  Johnson]{escobedo2025}
F.~J. Escobedo, K.~Yadav, O.~Cappelluti, and N.~Johnson, ``Exploring urban
  vegetation type and defensible space’s role in building loss during
  wildfire-driven events in california,'' \emph{Landscape and Urban Planning},
  vol. 262, p. 105421, 2025.

\bibitem[Kenny et~al.(2026)Kenny, Johns, Pawlak, Fricker, Yost, and
  Ritter]{kenny2026}
R.~Kenny, J.~Johns, C.~Pawlak, A.~Fricker, J.~Yost, and M.~Ritter, ``Urban
  trees and structure loss in the 2025 eaton and palisades fires,'' \emph{Urban
  Forestry \&amp; Urban Greening}, vol. 121, p. 129470, 2026.

\bibitem[Narimani et~al.(2025{\natexlab{a}})Narimani, Pourreza, Moghimi,
  Farajpoor, Jafarbiglu, and Mesgaran]{narimani2025b}
M.~Narimani, A.~Pourreza, A.~Moghimi, P.~Farajpoor, H.~Jafarbiglu, and M.~B.
  Mesgaran, ``Early detection of branched broomrape ({Phelipanche ramosa})
  infestation in tomato crops by using leaf spectral analysis and machine
  learning,'' \emph{IFAC-PapersOnLine}, vol.~59, no.~23, pp. 114--119, 2025.

\bibitem[Batty(2013)]{batty2013}
M.~Batty, \emph{The New Science of Cities}.\hskip 1em plus 0.5em minus
  0.4em\relax Cambridge, MA: MIT Press, 2013.

\bibitem[Shi et~al.(2022)Shi, Goodchild, Batty, Li, Liu, Zhang,
  et~al.]{shi2022}
W.~Shi, M.~F. Goodchild, M.~Batty, Q.~Li, X.~Liu, A.~Zhang \emph{et~al.},
  ``Prospective for urban informatics,'' \emph{Urban Informatics}, vol.~1,
  p.~2, 2022.

\bibitem[Pradeep(2026)]{pradeep2026}
R.~M.~M. Pradeep, ``Integrating science and policy in a {HydroGIS} framework
  for urban water resilience,'' \emph{Frontiers in Sustainability}, vol.~7, p.
  1893462, 2026.

\bibitem[Narimani et~al.(2026{\natexlab{a}})Narimani, Mitra, and
  Farajpoor]{narimani2026b}
M.~Narimani, S.~Mitra, and P.~Farajpoor, ``From crown candidates to
  neighborhood screening: Integrating optical {GeoAI} and spatial modeling for
  urban-canopy assessment in davis, california,'' 2026, arXiv preprint.

\bibitem[Narimani et~al.(2026{\natexlab{b}})Narimani, Pourreza, Moghimi, and
  Farajpoor]{narimani2026d}
M.~Narimani, A.~Pourreza, A.~Moghimi, and P.~Farajpoor, ``{Sentinel-2} for crop
  yield estimation: A systematic review,'' \emph{Smart Agricultural
  Technology}, vol.~14, p. 102405, 2026.

\bibitem[Norlen et~al.(2026)Norlen, Sharma, and Escobedo]{norlen2026}
C.~A. Norlen, S.~Sharma, and F.~J. Escobedo, ``Socio-ecological impacts of the
  2025 los angeles urban fires on communities, neighborhoods, and homes,''
  \emph{Nature Communications}, vol.~17, no.~1, p. 3941, 2026.

\bibitem[Roberts et~al.(2017)Roberts, Bahn, Ciuti, Boyce, Elith,
  Guillera‐Arroita, Hauenstein, Lahoz‐Monfort, Schröder, Thuiller, Warton,
  Wintle, Hartig, and Dormann]{roberts2017}
D.~R. Roberts, V.~Bahn, S.~Ciuti, M.~S. Boyce, J.~Elith, G.~Guillera‐Arroita,
  S.~Hauenstein, J.~J. Lahoz‐Monfort, B.~Schröder, W.~Thuiller, D.~I.
  Warton, B.~A. Wintle, F.~Hartig, and C.~F. Dormann, ``Cross‐validation
  strategies for data with temporal, spatial, hierarchical, or phylogenetic
  structure,'' \emph{Ecography}, vol.~40, no.~8, pp. 913--929, 2017.

\bibitem[Valavi et~al.(2019)Valavi, Elith, Lahoz-Monfort, and
  Guillera-Arroita]{valavi2019}
R.~Valavi, J.~Elith, J.~J. Lahoz-Monfort, and G.~Guillera-Arroita, ``block{CV}:
  An {R} package for generating spatially or environmentally separated folds
  for k-fold cross-validation of species distribution models,'' \emph{Methods
  in Ecology and Evolution}, vol.~10, no.~2, pp. 225--232, 2019.

\bibitem[Ploton et~al.(2020)Ploton, Mortier, Réjou-Méchain, Barbier, Picard,
  Rossi, Dormann, Cornu, Viennois, Bayol, Lyapustin, Gourlet-Fleury, and
  Pélissier]{ploton2020}
P.~Ploton, F.~Mortier, M.~Réjou-Méchain, N.~Barbier, N.~Picard, V.~Rossi,
  C.~Dormann, G.~Cornu, G.~Viennois, N.~Bayol, A.~Lyapustin, S.~Gourlet-Fleury,
  and R.~Pélissier, ``Spatial validation reveals poor predictive performance
  of large-scale ecological mapping models,'' \emph{Nature Communications},
  vol.~11, no.~1, p. 4540, 2020.

\bibitem[{CAL FIRE}(2025{\natexlab{a}})]{calfire2025incident}
{CAL FIRE}, ``Palisades fire incident information,''
  \url{https://www.fire.ca.gov/incidents/2025/1/7/palisades-fire/}, 2025,
  california Department of Forestry and Fire Protection. Accessed 2026-08-12.

\bibitem[{National Interagency Fire Center}(2025)]{nifc2025wfigs}
{National Interagency Fire Center}, ``{WFIGS} interagency fire perimeters,''
  \url{https://data-nifc.opendata.arcgis.com/datasets/nifc::wfigs-interagency-fire-perimeters/about},
  2025, accessed 2026-08-12.

\bibitem[Muñoz-Sabater et~al.(2021)Muñoz-Sabater, Dutra, Agustí-Panareda,
  Albergel, Arduini, Balsamo, Boussetta, Choulga, Harrigan, Hersbach, Martens,
  Miralles, Piles, Rodríguez-Fernández, Zsoter, Buontempo, and
  Thépaut]{munozsabater2021}
J.~Muñoz-Sabater, E.~Dutra, A.~Agustí-Panareda, C.~Albergel, G.~Arduini,
  G.~Balsamo, S.~Boussetta, M.~Choulga, S.~Harrigan, H.~Hersbach, B.~Martens,
  D.~G. Miralles, M.~Piles, N.~J. Rodríguez-Fernández, E.~Zsoter,
  C.~Buontempo, and J.-N. Thépaut, ``Era5-land: a state-of-the-art global
  reanalysis dataset for land applications,'' \emph{Earth System Science Data},
  vol.~13, no.~9, pp. 4349--4383, 2021.

\bibitem[Guzman‐Morales and Gershunov(2019)]{guzman2019}
J.~Guzman‐Morales and A.~Gershunov, ``Climate change suppresses santa ana
  winds of southern california and sharpens their seasonality,''
  \emph{Geophysical Research Letters}, vol.~46, no.~5, pp. 2772--2780, 2019.

\bibitem[Keeley and Syphard(2019)]{keeley2019}
J.~E. Keeley and A.~D. Syphard, ``Twenty-first century california, usa,
  wildfires: fuel-dominated vs. wind-dominated fires,'' \emph{Fire Ecology},
  vol.~15, no.~1, p.~24, 2019.

\bibitem[{CAL FIRE}(2025{\natexlab{b}})]{calfire2025dins}
{CAL FIRE}, ``{CAL FIRE} damage inspection ({DINS}) data,''
  \url{https://data.ca.gov/dataset/cal-fire-damage-inspection-dins-data}, 2025,
  california Open Data Portal. Accessed 2026-08-12.

\bibitem[{Insurance Institute for Business \& Home Safety}(2025)]{ibhs2025}
{Insurance Institute for Business \& Home Safety}, ``The 2025 {LA}
  conflagrations,'' IBHS, Tech. Rep., 2025, technical field-study report.

\bibitem[Narimani et~al.(2025{\natexlab{b}})Narimani, Pourreza, Moghimi,
  Farajpoor, Jafarbiglu, and Mesgaran]{narimani2025a}
M.~Narimani, A.~Pourreza, A.~Moghimi, P.~Farajpoor, H.~Jafarbiglu, and M.~B.
  Mesgaran, ``Branched broomrape detection in tomato farms using satellite
  imagery and time-series analysis,'' in \emph{Proc. {SPIE}}, vol. 13475, 2025,
  p. 134750U.

\bibitem[Gorelick et~al.(2017)Gorelick, Hancher, Dixon, Ilyushchenko, Thau, and
  Moore]{gorelick2017}
N.~Gorelick, M.~Hancher, M.~Dixon, S.~Ilyushchenko, D.~Thau, and R.~Moore,
  ``Google earth engine: Planetary-scale geospatial analysis for everyone,''
  \emph{Remote Sensing of Environment}, vol. 202, pp. 18--27, 2017.

\bibitem[Farajpoor et~al.(2025{\natexlab{a}})Farajpoor, Pourreza, Narimani,
  El-Kereamy, and Fidelibus]{farajpoor2025c}
P.~Farajpoor, A.~Pourreza, M.~Narimani, A.~El-Kereamy, and M.~W. Fidelibus,
  ``Multi-trait spectral modeling for estimating grapevine leaf traits and
  nutrients,'' \emph{Plant Phenomics}, vol.~7, p. 100142, 2025.

\bibitem[Chakraborty et~al.(2025)Chakraborty, Pourreza, Peanusaha, Farajpoor,
  Khalsa, and Brown]{chakraborty2025}
M.~Chakraborty, A.~Pourreza, S.~Peanusaha, P.~Farajpoor, S.~D.~S. Khalsa, and
  P.~H. Brown, ``Integrating hyperspectral radiative transfer modeling and
  machine learning for enhanced nitrogen sensing in almond leaves,''
  \emph{Computers and Electronics in Agriculture}, vol. 234, p. 110195, 2025.

\bibitem[Scott and Burgan(2005)]{scott2005}
J.~H. Scott and R.~E. Burgan, ``Standard fire behavior fuel models: A
  comprehensive set for use with {Rothermel}'s surface fire spread model,''
  USDA Forest Service, Rocky Mountain Research Station, Fort Collins, CO,
  General Technical Report RMRS-GTR-153, 2005.

\bibitem[Rollins(2009)]{rollins2009}
M.~G. Rollins, ``Landfire: a nationally consistent vegetation, wildland fire,
  and fuel assessment,'' \emph{International Journal of Wildland Fire},
  vol.~18, no.~3, pp. 235--249, 2009.

\bibitem[{LANDFIRE}(2025)]{landfire2024}
{LANDFIRE}, ``{LANDFIRE} 2024 update ({LF} 2024): 40 {Scott} and {Burgan} fire
  behavior fuel models and forest canopy cover,'' \url{https://landfire.gov},
  2025, u.S. Geological Survey and USDA Forest Service. Accessed 2026-08-12.

\bibitem[{OpenStreetMap contributors}(2025)]{osm2025}
{OpenStreetMap contributors}, ``Openstreetmap database snapshot (2025-01-01),''
  \url{https://www.openstreetmap.org}, 2025, open Database License (ODbL).
  Historical snapshot retrieved via Overpass API, accessed 2026-08-12.

\bibitem[Flanagan et~al.(2011)Flanagan, Gregory, Hallisey, Heitgerd, and
  Lewis]{flanagan2011}
B.~E. Flanagan, E.~W. Gregory, E.~J. Hallisey, J.~L. Heitgerd, and B.~Lewis,
  ``A social vulnerability index for disaster management,'' \emph{Journal of
  Homeland Security and Emergency Management}, vol.~8, no.~1, p.
  0000102202154773551792, 2011.

\bibitem[{Centers for Disease Control and Prevention / Agency for Toxic
  Substances and Disease Registry}(2024)]{cdcsvi2022}
{Centers for Disease Control and Prevention / Agency for Toxic Substances and
  Disease Registry}, ``{CDC/ATSDR} social vulnerability index 2022 database,
  california,'' \url{https://www.atsdr.cdc.gov/place-health/php/svi/}, 2024,
  accessed 2026-08-12.

\bibitem[Farajpoor et~al.(2024)Farajpoor, Pourreza, Fidelibus, and
  El-Kereamy]{farajpoor2024}
P.~Farajpoor, A.~Pourreza, M.~W. Fidelibus, and A.~El-Kereamy, ``Multi-trait
  modeling for grape nutrient retrieval by proximal spectral sensing,'' in
  \emph{Autonomous Air and Ground Sensing Systems for Agricultural Optimization
  and Phenotyping {IX}}, ser. Proc. {SPIE}, vol. 13053, 2024.

\bibitem[Chen and Guestrin(2016)]{chen2016}
T.~Chen and C.~Guestrin, ``Xgboost,'' \emph{Proceedings of the 22nd ACM SIGKDD
  International Conference on Knowledge Discovery and Data Mining}, pp.
  785--794, 2016.

\bibitem[Lundberg and Lee(2017)]{lundberg2017}
S.~M. Lundberg and S.-I. Lee, ``A unified approach to interpreting model
  predictions,'' in \emph{Advances in Neural Information Processing Systems 30
  (NIPS 2017)}, Long Beach, CA, 2017, pp. 4765--4774.

\bibitem[Farajpoor et~al.(2025{\natexlab{b}})Farajpoor, Pourreza, and
  Narimani]{farajpoor2025a}
P.~Farajpoor, A.~Pourreza, and M.~Narimani, ``Domain adaptation approaches to
  improve leaf trait estimation by reducing error propagation in hybrid
  modeling,'' in \emph{{AGU} Fall Meeting Abstracts}, 2025, pp. GC33K--0929.

\bibitem[Saito and Rehmsmeier(2015)]{saito2015}
T.~Saito and M.~Rehmsmeier, ``The precision-recall plot is more informative
  than the roc plot when evaluating binary classifiers on imbalanced
  datasets,'' \emph{PLOS ONE}, vol.~10, no.~3, p. e0118432, 2015.

\bibitem[Key and Benson(2006)]{key2006}
C.~H. Key and N.~C. Benson, ``Landscape assessment ({LA}): Sampling and
  analysis methods,'' USDA Forest Service, Rocky Mountain Research Station,
  Fort Collins, CO, General Technical Report RMRS-GTR-164-CD, 2006, in:
  FIREMON: Fire Effects Monitoring and Inventory System.

\bibitem[Miller and Thode(2007)]{miller2007}
J.~D. Miller and A.~E. Thode, ``Quantifying burn severity in a heterogeneous
  landscape with a relative version of the delta normalized burn ratio
  (dnbr),'' \emph{Remote Sensing of Environment}, vol. 109, no.~1, pp. 66--80,
  2007.

\bibitem[Farajpoor and Narimani(2026)]{palisades2026data}
P.~Farajpoor and M.~Narimani, ``Palisades urban wildfire {GeoAI}:
  Structure-level predictors and spatially validated model outputs for the 2025
  palisades fire,'' 2026.

\bibitem[Davies et~al.(2018)Davies, Haugo, Robertson, and Levin]{davies2018}
I.~P. Davies, R.~D. Haugo, J.~C. Robertson, and P.~S. Levin, ``The unequal
  vulnerability of communities of color to wildfire,'' \emph{PLOS ONE},
  vol.~13, no.~11, p. e0205825, 2018.

\bibitem[Farajpoor et~al.(2025{\natexlab{c}})Farajpoor, Pourreza, Narimani,
  El-Kereamy, and Fidelibus]{farajpoor2025b}
P.~Farajpoor, A.~Pourreza, M.~Narimani, A.~El-Kereamy, and M.~W. Fidelibus,
  ``Leaf spectral reflectance prediction using multihead attention neural
  networks,'' in \emph{Proc. {SPIE}}, vol. 13475, 2025, p. 134750V.

\bibitem[Narimani et~al.(2026{\natexlab{c}})Narimani, Pourreza, and
  Farajpoor]{narimani2026c}
M.~Narimani, A.~Pourreza, and P.~Farajpoor, ``Mapping tomato cropping systems
  in california using {AlphaEarth} geospatial embeddings and deep learning
  analysis,'' 2026, arXiv preprint.

\bibitem[Narimani et~al.(2026{\natexlab{d}})Narimani, Anand, and
  Farajpoor]{narimani2026a}
M.~Narimani, V.~Anand, and P.~Farajpoor, ``Farmland extent and visible boundary
  mapping from 1 m {NAIP} imagery using residual {U-Net} and text-prompted
  {SAM} 3 refinement,'' 2026, arXiv preprint.

\bibitem[Zamanialaei et~al.(2025)Zamanialaei, San~Martin, Theodori, Purnomo,
  Tohidi, Lautenberger, Qin, Trouvé, and Gollner]{zamanialaei2025}
M.~Zamanialaei, D.~San~Martin, M.~Theodori, D.~M.~J. Purnomo, A.~Tohidi,
  C.~Lautenberger, Y.~Qin, A.~Trouvé, and M.~Gollner, ``Fire risk to
  structures in california’s wildland-urban interface,'' \emph{Nature
  Communications}, vol.~16, no.~1, p. 8041, 2025.

\bibitem[Narimani et~al.(2024)Narimani, Pourreza, Moghimi, Mesgaran, Farajpoor,
  and Jafarbiglu]{narimani2024}
M.~Narimani, A.~Pourreza, A.~Moghimi, M.~B. Mesgaran, P.~Farajpoor, and
  H.~Jafarbiglu, ``Drone-based multispectral imaging and deep learning for
  timely detection of branched broomrape in tomato farms,'' in \emph{Proc.
  {SPIE}}, vol. 13053, 2024, p. 1305304.

\end{thebibliography}

\end{document}